\documentclass[prd,aps,10pt,reprint,nofootinbib,superscriptaddress,floatfix,onecolumn]{revtex4-2}

\usepackage{graphicx}
\usepackage{dcolumn}
\usepackage{bm}
\usepackage{booktabs}
\usepackage{amsmath,amssymb}
\usepackage{tensor}
\usepackage{titlesec}
\usepackage{xcolor}
\usepackage{ytableau}
\usepackage[colorlinks,linkcolor=blue]{hyperref}
\usepackage{ytableau}
\usepackage{orcidlink}
\usepackage{academicons}
\usepackage{mathrsfs}
\usepackage{caption}
\usepackage{subcaption}
\usepackage{tikz}
\usetikzlibrary{arrows.meta,calc,positioning}

\usepackage{todonotes}
\usepackage{marginnote}

\usepackage[normalem]{ulem}
\usepackage{cancel}

\definecolor{orcidlogocol}{HTML}{A6CE39}
\newcommand{\orcid}[1]{\href{https://orcid.org/#1}{\textcolor[HTML]{A6CE39}{\aiOrcid}}}

\def\al{\alpha}
\def\be{\beta}
\def\ga{\gamma}
\def\de{\delta}
\def\ep{\epsilon}
\def\ze{\zeta}
\def\et{\eta}
\def\th{\theta}
\def\ka{\kappa}
\def\la{\lambda}

\def\vp{\varphi}
\def\si{\sigma}
\def\ta{\tau}

\def\vph{\varphi}
\def\ch{\chi}
\def\ps{\psi}
\def\om{\omega}
\def\Ga{\Gamma}
\def\De{\Delta}
\def\Th{\Theta}

\def\Si{\Sigma}

\def\Om{\Omega}

\def\hvp{{\hat \vp}}
\def\hpi{{\hat \pi}}

\def\mn{{\mu\nu}}

\def\prt{\partial}
\def\pt#1{\phantom{#1}}

\def\cR{{\cal R}}
\def\rf#1{(\ref{#1})}
\def\bp{{\bar p}}

\def\tz{{\tilde z}}
\def\tac{{\tilde a}}
\def\tap{{\tau^\prime}}

\newcommand{\beq}{\begin{equation}}
\newcommand{\eeq}{\end{equation}}
\newcommand{\bal}{\begin{aligned}}
\newcommand{\eal}{\end{aligned}}

\usepackage{etoolbox}

\makeatletter
\patchcmd{\frontmatter@abstract@produce}
  {\vskip200\p@\@plus1fil
   \penalty-200\relax
   \vskip-200\p@\@plus-1fil}
  {}
  {}
  {}
\makeatother

\allowdisplaybreaks

\begin{document}

\title{Accelerating systems as probes of Lorentz violation}

\author{Quentin G.\ Bailey}
\email{baileyq@erau.edu}
\affiliation{Department of Physics and Astronomy, Embry-Riddle Aeronautical University, 3700 Willow Creek Road, Prescott, AZ, USA}
\author{Nils A. Nilsson}
\email{nilsson@ibs.re.kr}
\affiliation{Cosmology, Gravity and Astroparticle Physics Group, Center for Theoretical Physics of the Universe, Institute for Basic Science, Daejeon 34126, Korea}
\affiliation{LTE, Observatoire de Paris, Université PSL, CNRS, LNE, Sorbonne Universit\'e, 61 avenue de l’Observatoire, 75 014 Paris, France}
\author{Sawyer J.\ Star}

\affiliation{Department of Physics and Astronomy, Embry-Riddle Aeronautical University, 3700 Willow Creek Road, Prescott, AZ, USA}
\affiliation{Department of Physics and Astronomy, Washington State University, 1245 Webster, Pullman, WA, USA}

\date{\today}

\begin{abstract}

We investigate the Fulling-Davies-Unruh effect in a local Lorentz-violating effective field theory, focusing on the quantum-field response of accelerated detectors. We canonically quantize a real scalar field exactly in the coefficients for Lorentz violation.  The dispersion relation, Hamiltonian, and massless and massive Wightman functions are obtained.  
We determine the conditions under which the scalar dynamics preserves boost symmetry and distinguish these from the weaker conditions required for stationarity along a single accelerated worldline.
The Wightman functions are used to construct the finite-time Unruh-DeWitt response for a uniformly accelerated detector. 
For generic coefficients the response is nonstationary and depends on both the duration and temporal location of the measurement.  
The spectrum, 
which is the  resulting excitation rate as a function of the energy level difference in the detector, 
is obtained numerically.  
It shows discernable differences with the usual massive thermal spectrum for dimensionless coefficients on the order of $10^{-7}$.
Experimental implications of the result are discussed.
As an alternative viewpoint, we place Lorentz violation in the point-particle sector, derive the exact trajectory generated by a constant electric field, 
and study the proper-time Fourier response of a conventional scalar wave along the modified trajectory.
\end{abstract}


\maketitle

\section{Introduction}\label{sec:intro}

Presently, the physics of accelerated quantum systems is of interest from several directions, ranging from the foundations of quantum field theory to proposals for observing acceleration-induced effects in the laboratory. The best known result in this subject is the Fulling-Davies-Unruh (FDU) effect \cite{Fulling:1972md,Davies:1974th,unruh76}, according to which the Minkowski vacuum has a thermal character for a uniformly accelerated observer. For a boost orbit with proper acceleration $a$, the associated temperature is $T=a/(2\pi)$, and the same statement can be formulated in terms of the Kubo--Martin--Schwinger condition for the vacuum restricted to a Rindler wedge. The FDU effect is closely connected to the general observer dependence of the particle concept in quantum field theory and to Hawking radiation, and has been reviewed from a number of viewpoints \cite{Crispino2008,Frodden_2018,birrell_davies_1982,Takagi:1986kn}. 

A useful way to discuss the effect is through an Unruh--DeWitt detector, where a localized quantum system is carried along an accelerated worldline and its excitation probability is calculated from the field correlation function sampled on that trajectory.
This effect is however not simple to detect experimentally. Uniform acceleration for an arbitrarily long proper time is not available, and in any realistic setup the acceleration has to be produced by an interaction acting on the detector or on the particles used as probes. High-energy channeling experiments have provided an interesting setting in this regard. Evidence interpreted as acceleration-induced thermality and related horizon effects has been reported using channeling-radiation data \cite{Lynch:2019hmk,Lynch:2019xfl,Lynch:2023zll,Lynch:2025igp}, and further experimental proposals continue to be developed \cite{Janson:2026tqd}. The interpretation of such observations as a direct measurement of the standard Unruh effect is not settled, however, and theoretical objections to some detector models used in this context have also been raised \cite{Levin:2025thermality}. In addition, a detector can only interact with the field for a finite interval of proper time. The finite-time response is therefore a relevant observable in its own right, particularly when the background theory does not possess the stationarity that is normally used to pass to an eternal-detector transition rate.

At the same time, 
searches for small violations of spacetime symmetries remain a broad program in theory and experiment \cite{datatables,Safronova:2017xyt,Addazi:2021xuf,Mariz:2022oib}. A convenient theory-agnostic description is provided by effective field theory, in which conventional matter and field actions are supplemented by operators contracted with coefficients for Lorentz and CPT violation \cite{ck97,ck98}. In the context of accelerated detectors, Lorentz violation is particularly interesting because the standard FDU construction depends rather strongly on the relation between inertial time translations, boost evolution, and the causal structure of the quantum field.  
Once the field is coupled to a background tensor, 
these structures do not in general coincide with those of the Minkowski metric.

A number of approaches have been used to study this question. Modified dispersion relations, polymer quantization, and other field theories have been considered for accelerated and rotating detectors \cite{Kajuri:2015oza,Campo:2010fz,Hossain:2015xqa,Husain:2015tna,Carballo-Rubio:2018zll,Davies:2023zfq,Gutti:2010nv,DelPorro:2023knh,Xu:2025smb}. In polymer-quantized models, for example, the Kubo–Martin–Schwinger (KMS) property along a Rindler trajectory can be violated \cite{Hossain:2015xqa}, while modified-dispersion analyses show that the response can depend sensitively on whether the relevant modes are subluminal or superluminal \cite{Husain:2015tna,Davies:2023zfq,Gutti:2010nv}. 
In low-energy Lorentz-violating gravity, suitably defined accelerated configurations can nevertheless retain Unruh-like properties \cite{DelPorro:2023knh}. More recently, detector coherence and momentum-resolved detector observables have also been proposed as probes of Lorentz violation \cite{Xu:2025smb,Wu:2026phx}. 
Furthermore, studies of particle detectors in the presence of black holes with Lorentz violation exist~\cite{Liu:2025bpp}. 
These results show that accelerated-detector physics can be sensitive to symmetry-breaking effects even when the Lorentz-violating corrections are small.

There is a second issue which is important for the problem considered here. Much of the discussion in the literature of Lorentz violation and the FDU effect begins by specifying a modified dispersion relation and then asking how an accelerated detector responds. In an effective-field-theory setting, however, one may instead start directly from the local action. This permits the scalar theory to be quantized canonically, fixes the positive-frequency branch and vacuum used in the calculation, and gives the two-point function without introducing a perturbative expansion in the Lorentz-violating coefficients. It also makes clear which part of the result follows from the modified field correlations and which part follows from the accelerated trajectory used to sample them
\footnote{See Refs.\ \cite{km09,km13,Guerrero:2025awe} for a discussion of the effective field theory (EFT) approach to describing spacetime-symmetry violations compared with other approaches.}.

Finite interaction times provide another reason to keep this distinction explicit. It is well known that a detector operated for a finite time contains switching and transient effects which are absent from the idealized eternal-detector rate \cite{Sriramkumar:1996finite,Louko:2008transition,Stargen:2026finite}. In a stationary theory these effects can be separated from the long-time response under suitable conditions. When Lorentz violation destroys stationarity along the accelerated trajectory, this simplification is no longer available: the detector response can depend not only on the duration of the measurement but also on when the measurement is performed relative to the preferred frame. Recent work in which the detector's center-of-mass motion is treated dynamically also emphasizes that the motion of the probe and the response of its internal degrees of freedom need not carry identical thermal information \cite{Stargen:2025thermality}.

The paper is organized as follows. In Section~\ref{eft}, we introduce the Lorentz-violating point-particle, scalar, and detector sectors and discuss the two complementary coefficient assignments used in the calculations. 
In Section~\ref{sec:scalarfield_quantisation}, we canonically quantize the scalar field and derive its exact massless and massive Wightman functions. Section~\ref{sec:UnruDavies_boboliubov} discusses boost symmetry and the conditions for stationarity along an accelerated worldline. We then construct the finite-time Unruh--DeWitt response, including the dependence on the center of the measurement window and the weak boost-breaking expansion. In Section~\ref{detector}, we present the numerical detector response. In Section~\ref{FTM}, we turn to the complementary point-particle description, derive the modified accelerated trajectory, and use a proper-time Fourier transform of a scalar plane wave as an alternative diagnostic of the modified motion. Finally, in Section~\ref{sec:conclusions}, we summarize the results and discuss possible extensions. We use units in which $c=\hbar=k_B=1$, and the spacetime metric has signature $(-+++)$.

\section{Effective Field Theory with Lorentz violation}
\label{eft}

In this work, 
we use an action-based framework\footnote{We point out that the word ``framework'' has been used in this context long before the era of AI.} to describe deviations from perfect global Lorentz invariance (we assume flat spacetime in this work).
The structure of this framework posits
conventional actions for known matter and fields supplemented with symmetry-breaking
terms controlled by ``coefficients" for CPT and Lorentz violation \cite{kp95,ck97,ck98}.
The action relevant for this work takes the form
\beq
S=S_{pp}+S_{\vph}+S_{Int}+...,
\label{action}
\eeq
where $S_{pp}$ is the point-particle action, $S_{\phi}$ is the scalar action, 
and $S_{Int}$ is the interaction term between the scalar and the point particles.
The ellipses include other possible terms, 
for instance, 
a conventional electromagnetic coupling to mimic a constant acceleration.
Also note that rather than study the FDU effect with electromagnetic fields, 
we use scalars for simplicity in this initial investigation.

To model an accelerating particle affected by Lorentz violation we use the following action:
\beq
S_{pp} = -m \int \sqrt{(\et_\mn + c_\mn) \dot{x}^\mu \dot{x}^\nu } d\la
\label{particle}
\eeq
where $\dot{x}^\mu = dx^\mu/d\la$ \cite{kl10,kt11}.
Here the coefficients for Lorentz violation are $c_\mn$, 
which is symmetric and contains $10$ {\it a priori} unknown quantities 
that describe the degree of Lorentz violation for a given type of particle.  
These coefficients indeed could depend on the type of particle 
and break the Weak-Equivalence Principle (WEP) as well.
We assume that in inertial cartesian coordinates the coefficients are constants
$\prt_\mu c_{\nu\la}=0$.

The action for the scalar field we consider is 
\begin{equation}
    S_{\vp}=-\frac{1}{2}\int d^4x \left[(\eta^{\mu\nu}+k^{\mu\nu})\partial_\mu\varphi \partial_\nu \varphi + m^2 \varphi^2 \right],
    \label{scalar_act}
\end{equation}
where $m\geq0$ is the mass and the $k^\mn$ are the coefficients for spacetime-symmetry breaking \cite{ck98}.
It is useful to define the quantity $K^{\mu\nu}$ as
\begin{equation}
	K^{\mu\nu}\equiv \eta^{\mu\nu}+k^{\mu\nu},
    \label{Kdef}
\end{equation}
where we assume that $k^{\mu\nu}$ is a constant in the chosen coordinates and that $K^{\mu\nu}$ has Lorentzian signature. Also, in order for $t\equiv x^0$ to be a well-defined time evolution parameter we must have that
\begin{equation}
    K^{00}<0, \qquad K^{ij}\,\,\text{positive definite},
    \label{Kinv}
\end{equation}
which also guarantees hyperbolicity and positivity of the Hamiltonian in some special coordinates. 
Note that $K^{\mu\nu}$ does not need to be diagonal or pertubatively small. 
These are the only {\it a priori} constraints on $K^{\mu\nu}$; further limitations may be imposed from physical arguments where one expects that any Lorentz violation in nature should be small. We make no such assumption here.
In Sections \ref{sec:scalarfield_quantisation}-\ref{detector}, we place Lorentz violation in the scalar sector and use the Minkowski metric $\eta_{\mu\nu}$ to define the proper time and acceleration of the detector. In this case, scalar propagation is instead governed by the effective (or acoustic) metric $(K^{-1})_{\mu\nu}$. In particular, the characteristic surfaces satisfy $K^{\mu\nu}q_\mu q_\nu=0$, where $q_\mu$ is normal to the surface. The causal regions relevant for scalar signals are then determined by $(K^{-1})_{\mu\nu}$, which generally does not describe the same null cone as the Minkowski metric. Unless otherwise indicated, we raise and lower indices using the Minkowski metric $\eta^{\mu\nu}$.

Finally, 
the interaction term $S_{Int}$ is taken to describe a scalar field interacting with a two-level quantum system moving along the worldline of the point particle.
Specifically, 
\beq
S_{Int} = \int^{\infty}_{-\infty} \ch (\ta ) {\hat M} (\ta) {\hat \vph} [ x(\ta)] d\la
\label{Sint}
\eeq
where ${\hat M}$ is a Hermitian operator acting on a two-dimensional Hilbert space representing the detector quantum system, as detailed elsewhere \cite{dewitt1979quantum,Crispino2008}.
For our purposes we assume two energy levels $E_1$ and $E_2$ for this system.
The function $\ch$ is the detector response function, 
which controls how long the interaction is effectively ``on"
\cite{Salton_2015}.
An expansion to lowest order in pertubation theory can be used to find the simplest expression for the transition rate from the ground state to the excited state for a detector. 
Since we do not modify this interaction term directly with any coefficients for spacetime symmetry breaking, 
we can use some published results later in the paper.

Note that the coefficients $c_\mn$ and  $k_\mn$ for the point particles and the scalar field are identical in mathematical structure but could take on independent values if
their origin is a relic from an unspecified theory of quantum gravity.
They may or may not be related and we make no assumptions in this regard.
However, 
it has been shown in several works, 
that coefficients these types, that come paired with the Minkowksi metric $\et_\mn$ 
can be represented by an effective spacetime metric, 
like $K^\mn$ and $\et_\mn + c_\mn$. 
In fact, 
for the case of single particle species, 
one could employ a skew coordinate transformation to ``move"
the coefficients from one sector to another.
This process has been explained in detail elsewhere \cite{Bailey:2004na,kt11,Bailey:2023lzy}.
The physically measureable coefficients end up being a linear combination of the $k_\mn$ and $c_\mn$, 
representing a difference between two sectors.
One can then, 
equivalently, 
work with the coefficients in just one sector for calculations.
To find the result when the coefficients are in the other sector, 
one simply undoes the skew coordinate transformation
\cite{Yoder:2012ks}.

In the present paper, we will use two complementary descriptions. In the main quantum-field theory analysis we set the point-particle coefficients to zero and place the Lorentz-violating effects in the scalar sector with $k^{\mu\nu}\neq0$, so that the detector follows the standard uniformly accelerated trajectory while the vacuum correlations are modified. Later, we consider a complementary case $c_{\mu\nu}\neq0, \, k^{\mu\nu}=0$ in which the scalar field is conventional but the accelerated trajectory is modified.

\section{Exact quantization and Wightman functions}
\label{sec:scalarfield_quantisation}

We now quantize a real scalar field with constant coefficients for Lorentz-violation and thus constant $K^\mn$. 
Since this is constant in inertial Cartesian coordinates, 
translation invariance is preserved and the field admits a plane-wave expansion. 
Here, 
we present canonical results without resorting to leading-order approximations of the coefficients as in many treatments in the literature.
Note that some results to follow for the scalar in \rf{scalar_act} have been obtained in the relativistic quantum mechanics regime in Ref.\ \cite{Altschul:2022che}
and in other contexts \cite{Edwards:2018lsn,Furtado_2020,ONeal-Ault:2021uwu,Cabral:2026mxv}.
We also point out the quantization of a Lorentz-violating scalar field was studied in \cite{ArderucioCosta:2023vlj} for a one nonzero coefficient subset.
Variation of the action~\eqref{scalar_act} with respect to $\vp$ leads to the field equations,
\begin{equation}\label{eq:phieom}
\left[K^{\mu\nu}\partial_\mu\partial_\nu-m^2\right]\vp=0,
\end{equation}
where the first term contains the regular d'Alembertian; 
this is the massive Klein-Gordon equation with a $k^{\mu\nu}$ correction. 
In Cartesian coordinates, 
we can expand into time and space and obtain
	\begin{equation}
K^{00}\ddot{\vp}+2K^{0i}\partial_i\dot{\vp}+K^{ij}\partial_i\partial_j\vp-m^2\vp=0.
\end{equation}

We write the action \eqref{scalar_act} in 3+1 form as
	\begin{equation}
		S_\vp=-\frac{1}{2}\int d^4x\left[K^{00}\dot{\vp}^2+2K^{0i}\dot{\vp}\partial_i\vp+K^{ij}\partial_i\vp\partial_j\vp+m^2\vp^2\right],
	\end{equation}
and the Hamiltonian for the model is obtained by first finding the canonically conjugate momentum to $\varphi$ as $\pi_\vp (x) = \de \mathcal{L}/\de \dot {\vp}$ and we obtain,
\beq
\bal
\pi_\vp (x) = -K^{00}\dot{\vp}-K^{0i}\partial_i\vp .
\label{mom}
\eal
\eeq
Solving Eq.~\eqref{mom} for $\dot{\vp}$ and performing a Legendre transformation gives the Hamiltonian density as
	\begin{equation}
    \label{ham}
		\mathcal{H}=-\frac{1}{2K^{00}}\left(\pi_\vp+K^{0i}\partial_i\vp\right)^2+\frac{1}{2}K^{ij}\partial_i\vp\partial_j\vp+\frac{1}{2}m^2\vp^2.
	\end{equation}
	The first term is positive since $K^{00}<0$ which together with positive-definiteness of the spatial part of $K^{\mu\nu}$ makes the Hamiltonian density positive definite.
The resulting Hamilton equations are given by
\beq
\bal
\dot \vp &= -\frac{1}{K^{00}} (\pi_\vp + K^{0i} \prt_i \vp ), \\
\dot \pi_\vp &= -\frac{K^{0i}}{K^{00}}\prt_i (\pi_\vp+K^{0j}\prt_j \vp) 
+K^{ij}\prt_i \prt_j \vp-m^2 \vp.
\label{hamiltons}
\eal
\eeq
Since $K^{00}\neq0$, the relation between $\pi_\varphi$ and $\dot\varphi$ is invertible and the theory has no primary constraints. It therefore propagates one scalar degree of freedom.

\subsection{Dispersion relation and mode expansion}
\label{modes}

We first examine the field equations directly from the Euler-Lagrange equations \rf{eq:phieom}, 
rather than the Hamiltonian approach. For convenience, we define
\begin{equation}
    \ga :=
    -K^{00} >0,
    \label{eq:alpha}
\end{equation}
which we use henceforth. The field equation for $\vp$ is in Eq.~\eqref{eq:phieom},
which we decompose into Fourier modes using
\beq
\vp (x) = \frac{1}{(2\pi)^4} \int d^4p \, \tilde \vp (p) e^{i \, p \cdot x},
\label{Fourier}
\eeq
where $p\cdot x=p_\mu x^\mu$.
This yields the momentum space equations given by
\beq
\bal
K^{\mu\nu}p_\mu p_\nu +m^2 &=0,
\label{momspace}
\eal
\eeq
where the four-momentum is $p^\mu=(\om,\bf p)$.
For later use, 
we define
$f_\mathbf{p}(\om)=\ga \om^2 + 2 \om K^{0j} p_j-K^{jl}p_j p_l -m^2=0$.
The solutions to \rf{momspace}, or $f_\mathbf{p}(\om)=0$, 
are
\beq
\omega_\pm(\mathbf p)=\frac{1}{\ga}(\pm
\bar\omega_{\mathbf p}-K^{0i}p_i),
\quad
\bar\omega_{\mathbf p}
\equiv
\sqrt{
\left(K^{0i}p_i\right)^2
+
\ga\left(K^{ij}p_ip_j+m^2\right)
},
\label{disp}
\eeq
This dispersion relation shows anisotropic dependence on momentum direction $\hat p$.
Since $K^{ij}$ is positive definite and $\ga>0$, we have that
$
\bar\omega_{\mathbf p}^2
-
\left(K^{0i}p_i\right)^2
=
\ga\left(K^{ij}p_ip_j+m^2\right)>0
$. 
Therefore $\bar\omega_{\mathbf p}>|K^{0i}p_i|$ and hence 
$
\omega_+(\mathbf p)>0,\, \omega_-(\mathbf p)<0.
$
Nonetheless superluminal or subluminal speeds with respect to the Minkowski metric are possible since the causal cone for $\varphi$ is determined by $K^{\mu\nu}$.
However, this assumes light travels along the standard light cone, while it is well known that spacetime-symmetry breaking can modify the light cone for electromagnetic waves \cite{km02,Schreck:2011ai} independently.

We return now to the expansion in Eq.~\rf{Fourier} and insert a delta function with argument $f(\om)$ to enforce the dispersion relation. 
Thus we take $\tilde \vp = a(p) \de (f(\om) )$ and, 
\beq
\bal
\vp (x) &= \frac{1}{(2\pi)^4} \int d^4p \,  
a(p) \de (f(\om) )
e^{i \, p \cdot x},\\
&= 
\frac{1}{(2\pi)^4} \int 
\frac {d^3p}{2\bar{\om}_{\bf p}}
\left( 
a(\om_+,{\bf p} )
e^{-i ( \om_+ t - {\bf p} \cdot {\bf x})}
+a(\om_-,{\bf p})
e^{-i ( \om_- t - {\bf p} \cdot {\bf x})}
\right),
\label{Fourier2}
\eal
\eeq
where the second line is obtained by breaking up the delta function into two terms with the two roots \rf{disp}.
In the second term,
we re-write the momentum integral using $\omega_-(-\mathbf{p})=-\omega_+(\mathbf{p})$ followed by taking $\bf p \rightarrow -\bf p$ 
and then, 
upon demanding $\vp$ be real valued, 
the field expansion can be written in terms of $\om_+$ only:
\beq
\vp (x)=
\frac{1}{(2\pi)^4} \int \frac {d^3p}{2\bar{\om}_{\bf p}}
\left( 
a_{\bf p} \,
e^{-i ( \om_+ t - \bf p \cdot \bf x)}
+a_{\bf p}^* \,
e^{i ( \om_+ t - \bf p \cdot \bf x)}
\right),
\label{Fourier3}
\eeq
where $a_{\bf p}=a(\om_+ ,\bf p )$.
From this expression and \rf{mom} we can also obtain the canonical momentum $\pi$ in a mode expansion:
\beq
\pi (x)=
\frac{-i}{(2\pi)^4} \int d^3p \, 
\frac 12
\left( 
a_{\bf p} \,
e^{-i ( \om_+ t - \bf p \cdot \bf x)}
-a_{\bf p}^* \,
e^{i ( \om_+ t - \bf p \cdot \bf x)}
\right).
\label{mom2}
\eeq

We now promote the mode expansion to an operator expansion and impose the canonical commutation relations.
The fields in \rf{Fourier3}-\rf{mom2} are promoted to operators (and thus $a_{\bf p} \rightarrow \hat{a}_{\bf p}$).
The canonical commutation relations are then imposed:
\beq
\bal
 \relax [ \hvp (t, {\bf x} ), \hpi (t,{\bf x^\prime} ) ] &= 
i \de^{(3)} ({\bf x} - {\bf x^\prime} ),   \\
[ \hvp (t, {\bf x} ), \hvp (t, {\bf x^\prime} ) ] &= 0,   \\
[ \hpi (t, {\bf }x ), \hpi (t,{\bf x^\prime} ) ] &= 0.
\label{ccr}
\eal
\eeq
Using these commutation relations, 
and an appropriate rescaling of $\hat a_{\bf p}$,
we can express the quantized field and conjugate momentum as
\beq
\bal
\hat {\vp} (x)
&= 
\frac{1}{(2\pi)^3} 
\int d^3p \, 
\frac {1}
{ \sqrt{2 {\bar \omega}_{\mathbf p} } }
\left( 
{\hat b}_{\bf p} \,
e^{-i ( \om_+ t - \bf p \cdot \bf x)}
+ {\hat b}_{\bf p}^\dagger \,
e^{i ( \om_+ t - \bf p \cdot \bf x)}
\right),
\\
\hat {\pi} (x)&=
-i \frac{1}{(2\pi)^3} 
\int d^3p \, 
\sqrt{\frac{\bar\omega_{\mathbf p}}{2}}
\left( 
{\hat b}_{\bf p} \,
e^{-i ( \om_+ t - \bf p \cdot \bf x)}
- {\hat b}_{\bf p}^\dagger \,
e^{i ( \om_+ t - \bf p \cdot \bf x)}
\right),
\label{fieldQ}
\eal
\eeq
where the creation and annihilation operators satisfy:
\beq
\bal
[\hat b_{\bf p}, \hat b^\dagger_{\bf p^{\prime}}]&=(2\pi)^3 \de^{(3)} (\bf p - \bf p^\prime ), \\
[\hat b_{\bf p}, \hat b_{\bf p^{\prime}}]&=0, \\
[\hat b_{\bf p}^\dagger, \hat b^\dagger_{\bf p^{\prime}}]&=0, \\
\label{bcr}
\eal
\eeq
and the vacuum is defined by $\hat b_{\bf p}|0\rangle=0$.

We return to the Hamiltonian \rf{ham},
and insert the quantized fields to calculate $\hat H$.
After a lengthy calculation, 
the result can be written as
\begin{equation}
    {\hat H} = \int \frac{d^3p}{(2\pi)^3} \, \omega_+ \left[ {\hat b}_{\bf p}^\dagger \, {\hat b}_{\bf p}\right].
\label{ham2}
\end{equation}
This expression takes a conventional form albeit with the modified energy in Eq.~\rf{disp} that depends on the coefficients $K^\mn$. However, since $\omega_+(\mathbf p)>0$, all single-particle excitations have positive Hamiltonian energy in the quantization frame. Furthermore, since the scalar causal cone need not coincide with the Minkowski light cone, an inertial time coordinate obtained by an arbitrary Minkowski boost need not remain an admissible Hamiltonian time for the scalar field. We therefore define positive frequency with respect to the Cartesian time $x^0$, for which the conditions $K^{00}<0$ and $K^{ij}>0$ imposed above ensure a well-defined Hamiltonian evolution. Positive-frequency modes are those proportional to
\begin{equation}
    e^{-i\omega_+x^0}, \quad \omega_+>0,
\end{equation}
and the associated vacuum is the state defined by
$\hat b_{\mathbf p}|0\rangle=0$.

One can also construct 
the canonical energy-momentum tensor from \rf{scalar_act}, 
using the usual formula
$\Th^\mu_{\pt{\mu}{\nu}} =  \prt_\nu \vph  \prt {\cal L}/\prt (\prt_\mu \vph)- \de^\mu_{\pt{\mu}\nu} {\cal L}$. 
It obeys $\prt_\mu \Th^\mu_{\pt{\mu}{\nu}}=0$, which is due to the translation invariance of the action.
This quantity can be used, 
together with \rf{fieldQ}, 
to verify \rf{ham2} and to construct the momentum operator ${\hat P}^j = \int d^3x \Th^{0j}$, 
the latter of which takes the conventional form:
\beq
{\hat P^j} =  \int \frac{d^3p}{(2\pi)^3} \, p^j \left[ {\hat b}_{\bf p}^\dagger \, {\hat b}_{\bf p}\right].
\label{momentum}
\eeq
Of course, 
because the framework \rf{scalar_act} explicitly breaks global Lorentz invariance, 
the energy-momentum tensor cannot be symmetrized without spoiling its conservation and the six conserved angular momentum quantities 
are not conserved. It is also not clear that the Belifante procedure as outlined in \cite{Saffer:2017ywl} will help. This is in fact expected (for example, $\Th^\mn \neq \Th^{\nu\mu}$).
For more discussion on this point see Ref.\ \cite{ck97} and \cite{k04}.

\subsection{Wightman functions}
\label{wightman functions}
It is useful for calculations that follow to find the two-point function for the scalar field.
These are often referred to as Wightman functions
\cite{wightman56}, 
which were computed for a one coefficient sample in Ref.\ \cite{ArderucioCosta:2023vlj}. 
These can be found from the vacuum expectation value of two fields at two distinct spacetime points $x$ and $y$:
\beq
W(x,y)=\langle 0 | \hvp (x) \hvp (y) | 0 \rangle .
\label{twopoint}
\eeq
Since the field is modified by the $k_\mn$ coefficients, we outline how to find the result in some detail.
Inserting the field expansions from \rf{fieldQ} into \rf{twopoint}, 
we readily obtain
\beq
W(z)= \frac{1}{(2\pi)^3} \int d^3 p \frac {1}{2 {\bar \omega}_{\mathbf p} } e^{-i (\om_+ z^0 - \bf p \cdot \bf z) },
\label{twopoint2}
\eeq
where $z^\mu =x^\mu -y^\mu$.
The reader should note that this result depends on the modified frequency $\om_+=({\bar \omega}_{\mathbf p} +k_{0j}p^j)/(1-k_{00})$ with the quantity ${\bar \omega}_{\mathbf p}$ defined in previously in section \ref{modes}.
The coefficients $k_\mn$ control anisotropic terms in the integral \rf{twopoint2}.\\[2mm]

\noindent\underline{\textit{Massless field}}\\

Consider first the case of $m=0$.
In this case, 
we re-write the integral \rf{twopoint2} in the suggestive form:
\beq
W(z)= \frac{1}{(2\pi)^3 \sqrt{\ga}}
\int d^3 p \frac {1}{2 \sqrt{p^j p^l 
(\de_{jl}+C_{jl})}} 
e^{-i \left[ 
(\sqrt{\ga p^j p^l (\de_{jl}+C_{jl})} 
+k_{0j}p^j)
\tfrac {z^0}{\ga} 
- \bf p \cdot \bf z \right] },
\label{twopointml}
\eeq
where $C_{jl}=k_{jl}+k_{0j} k_{0l}/\ga$,
and $\ga = 1-k_{00}$.
The matrix $I+C$ is positive definite since $K^{ij}$ is positive definite and $k_{0i}k_{0j}/\ga$ is positive semidefinite. 
To solve the integral, 
we apply a linear transformation $p^j =M^j_{\pt{j}k} \bar p^k$ such that
\beq
p^j p^l (\de_{jl} + C_{jl} ) = \bp^j \bp^l \de_{jl},
\label{lintrans}
\eeq
thus rendering the functions in the integral spherically symmetric in the new $\bp$ variables to simplify the integration.
Note that this implies the transformation $M^j_{\pt{j}k}$ satisfies the matrix equation $M^T \cdot (I+C) \cdot M=I$.
This type of method has been adopted before in Ref.\ \cite{Bailey:2023lzy}.
Taking the determinant of this relation gives $(\det M)^2\det(I+C)=1$,
and therefore
$|\det M|=1/\sqrt{\det(I+C)}$. Since $\mathbf p=M\bar{\mathbf p}$, the momentum-space measure transforms as $d^3p=|\det M|\,d^3\bar p= d^3\bar p/\sqrt{\det(I+C)}$, and the integral in \rf{twopointml} becomes
\beq
W(z)= \frac{1}{(2\pi)^3 \sqrt{\ga}} 
\frac {1}{\sqrt{|I+C|} } 
\int d^3 \bp \frac {1}{2 \bp } e^{-i (\bp \tilde {z}^0 - \bf \bp \cdot \bf \tz) },
\label{twopointml2}
\eeq
where $\tilde z^0
=z^0/\sqrt{\ga},\,
\tilde{\mathbf z}=M^T
(\mathbf z-z^0\mathbf k_0/\ga)$
The normalization factor can also be written directly in terms of the
kinetic tensor. Using that $K^{00}=-\ga$, we use the Schur complement identity \cite{Zizong2009} to write 
$\det K=-\gamma\,\det(I+C)$.
Consequently, we have that
\begin{equation}
\frac{1}{\sqrt{\ga}\sqrt{\det(I+C)}}
=
\frac{1}{\sqrt{-\det K}}.
\label{eq:Wightman_normalization_identity}
\end{equation}

This result \rf{twopointml2} now appears mathematically identical to the standard case but with an overall scaling and modified $z$ argument. In particular, we emphasize that $\bp=\sqrt{\bp^j \bp^k \de_{jk}}$.
To ensure convergence and specify how distributional singularities are approached, we introduce a small damping factor $e^{-\epsilon\bar{\omega}}$ for $\epsilon>0$ with $\epsilon\to0^+$. Here, $\bar{\omega}$ is the positive frequency after a linear change in momentum variables in the Wightman function. We can write the damping factor as
$e^{-i\bar{\omega}\tilde{z}^0}e^{-\epsilon\bar{\omega}}=e^{-i\bar{\omega}(\tilde{z}^0-i\epsilon)}$,
and therefore, 
the positive-frequency prescription is
\begin{equation}
    \tilde{z}^0 \, \to \, \tilde{z}^0-i\epsilon,
    \label{eq:preferred_frame_iepsilon}
\end{equation}
which guarantees that the positive-frequency branch with respect to the present inertial frame is regularized.
The integral can be evaluated using spherical coordinates,
$\bf \bp \cdot \bf \tz = \bp \, \tz \cos \th $, 
and the 
 result for the two-point function is
\begin{equation}
W^+_{0,\epsilon}(z)
=
\frac{1}{4\pi^2\sqrt{-\det K}}
\frac1{Q_\epsilon(z)},
\label{eq:general_massless_Wightman}
\end{equation}
where we have defined
\begin{equation}
Q_\epsilon(z)
\equiv
\widetilde{\mathbf z}^{\,2}
-
(\widetilde z^0-i\epsilon)^2,
\label{eq:Qepsilon_general}
\end{equation}
and where the exact positive Wightman function $W^+(z)$ is obtained by taking $\lim_{\epsilon\to0^+}W^+(z)$, as we show in Appendix~\ref{app:pullbackiepsilon}.
Note that the spacetime distance-like quantity in the denominator matches the modified null path found in Ref.\ \cite{Bailey:2023lzy} for a classical scalar field Green function.

As an example of the effect of the $k_\mn$ coefficients on Wightman function, consider the simplified case where the only nonzero coefficients are $k_{11}$ and $k_{00}$.
We further assume the the overall trace of $k_\mn$ vanishes, $k_\al^{\pt{\al}\al}=0$, 
as a nonzero value would merely rescale the conventional result.
This implies that $k_{00}=k_{11}$, 
and the Wightman function can be written as
\beq
W^+_{0,\epsilon}(z) = \frac{1}{4\pi^2} 
\sqrt{\ze} 
\frac {1}{ (z^0-i\ep)^2 - \ze(z^1)^2 
-(1-k_{11}) (\bf z_\perp)^2},
\label{twopointml4}
\eeq
where $\ze=(1-k_{11})/(1+k_{11})$.\\[2mm]

\noindent\underline{\textit{Massive field}}:\\

For the massive case, the method is very similar but the result is substantially different due to the nonzero mass $m$ (the massive scalar does not travel along a null path).
We can re-write the integral \rf{twopoint2} in the form
\beq
W^+_m(z)= \frac{1}{(2\pi)^3 \sqrt{\ga}}
\int d^3 p 
\frac {1}{2 \sqrt{p^j p^l (\de_{jl}+C_{jl})+m^2}} 
e^{-i \left[ 
\sqrt{p^j p^l (\de_{jl}+C_{jl})+m^2} 
\tfrac {z^0}{\sqrt{\ga}} 
- p^j \left(z^j - \frac {k_{0j}z^0}{\ga} \right) \right] }.
\label{twopointm}
\eeq
We next employ the same transformation as in the massless case above (see \rf{lintrans}).  The result can then be expressed as
\beq
W^+_m(z)= \frac{1}{(2\pi)^3} 
\frac {1}{\sqrt{-\det K} } 
\int d^3 \bp \frac {1}{2 \sqrt{\bp^2+m^2} } e^{-i (\sqrt{\bp^2+m^2} \tilde {z}^0 - \bf \bp \cdot \bf \tz) }, 
\label{twopointm2}
\eeq
with the same expressions for $\tilde z^0$ and $\tilde z^j$ as for the massless case.

The integral in \rf{twopointm2} can be done in spherical coordinates as before, 
but the remaining integral over $\bp$ is non-trivial, but it can be related to the modified Bessel function of the second kind $K_1(x)$ \cite{arfken,Tjoa:2021roz}. 
The result requires regularization with $\tilde{z}^0 \rightarrow \tilde{z}^0-i \ep$ as in the massless case, and yields

\beq
W^+_{m,\epsilon}(z) = \frac{1}{4\pi^2 \sqrt{-\det K} } 
\frac {mK_1 (m \sqrt{Q_\epsilon})}{\sqrt{Q_\epsilon}}.
\label{twopointm3}
\eeq
The branch of the square root is fixed by the
$i\epsilon$ prescription.

Equations~\eqref{eq:general_massless_Wightman} and
\eqref{twopointm3} give the exact inertial-frame
vacuom correlation functions for the scalar model. 
Their dependence on the separation is
controlled by the effective quadratic interval
$Q_\epsilon(z)$ associated with
$H_{\mu\nu}=(K^{-1})_{\mu\nu}$. For a detector following a worldline
$x^\mu(\tau)$, the response is obtained by evaluating this interval on
$z^\mu=x^\mu(\tau)-y^\mu(\tau')$. Whether the resulting pullback depends
only on $\tau-\tau'$ is determined by whether the accelerated detector flow
is a symmetry of the scalar effective geometry. We now determine the
complete class of kinetic tensors $K^{\mu\nu}$ for which this stationarity is preserved.

\section{Boost symmetry and the Bogoluibov transformation method
}
\label{sec:UnruDavies_boboliubov}

The standard derivation of the Unruh effect compares the quantized scalar field in the inertial frame \rf{fieldQ}
with the quantized scalar field in the accelerating observer frame.
The idea is that the creation and annhilation operators in both settings are unitarily inequivalent but are related by a linear transformation called a Bogoluibov transformation.
With such a relation, 
one can then calculate the vacuum expectation value of the particle number operator for the accelerating or Rindler observer and find that it is not zero and obeys a thermal distribution \cite{Frodden_2018}.
However, 
this standard method relies on the existence of a time-independent decomposition into modes of definite Rindler frequency. 
Such a decomposition is available only when the scalar-field operator is invariant under boosts.
Here, 
we first determine the complete class of kinetic tensors for which this condition holds. Outside this class, one may still introduce time-dependent mode bases, but no preferred stationary notion of Rindler particles exists.

From the $(t,\mathbf{x})$ coordinates, we introduce constant acceleration $a$ in the $\hat{x}^1$ direction, and we write the right Rindler coordinates as
	\begin{equation}
		t=\rho\sinh{\eta}, \quad x^1=\rho\cosh{\eta}, \quad \rho>0.
        \label{wl0}
	\end{equation}
    A trajectory at fixed $\rho$ has proper acceleration $a=1/\rho$ and proper time $\tau=\rho\eta$. Along the detector trajectory described by $\rho=1/a$, we therefore have $\eta=a\tau$. In these coordinates, the metric reads
    \begin{equation}
        ds^2=-\rho^2 d\eta^2+d\rho^2+dx_\perp^2,
    \end{equation}
	and we have the basis vectors,
$\partial_\rho=\sinh{\eta}\partial_t+\cosh{\eta}\partial_1, \partial_\eta=\rho(\cosh{\eta}\partial_t+\sinh{\eta}\partial_1)$,
	which can be obtained by the chain rule. 
    We can solve for the basis vectors in the inertial coordinates as
	\begin{equation}
		\partial_t=\frac{\cosh{\eta}}{\rho}\partial_\eta-\sinh{\eta}\partial_\rho, \quad \partial_1=-\frac{\sinh{\eta}}{\rho}\partial_\eta+\cosh{\eta}\partial_\rho.
	\end{equation}

Before considering the field equations \rf{eq:phieom},
we state the condition for Rindler stationarity in a covariant way. For acceleration in the $x^1$ direction, translations in Rindler time are generated by the timelike boost generator $\xi=\partial_\eta=x^1\partial_0+x^0\partial_1$.
We can write the scalar-field equations of motion~\eqref{eq:phieom} as $\mathcal{D}\varphi=0$, where $\mathcal{D}\equiv K^{\mu\nu}\partial_\mu\partial_\nu-m^2$ is the modified Klein-Gordon operator. The theory is symmetric under boosts as long as 
\begin{equation}
    [\mathcal{L}_\xi,\mathcal{D}]=0
\end{equation}
holds. Here, $\mathcal{L}_\xi$ is the Lie derivative along the generator $\xi^\mu$.
Since $K^{\mu\nu}$ is a constant in inertial Cartesian coordinates and $\xi^\mu$ is linear in the coordinates, the commutator acting on the field $\phi$ is
\begin{equation}
    [\mathcal{L}_\xi,\mathcal{D}]\varphi = (\mathcal{L}_\xi K)^{\mu\nu}\partial_\mu\partial_\nu\varphi.
\end{equation}

Since the mass term commutes with
$\mathcal L_\xi$, the operator-invariance condition applies
to both massive and massless fields. It also follows from
requiring invariance of all on-shell solutions. Plane waves require
$(\mathcal{L}_\xi K)^{\mu\nu}p_\mu p_\nu=0$ whenever
$K^{\mu\nu}p_\mu p_\nu=-m^2$. For $m>0$, this forces
$(\mathcal{L}_\xi K)^{\mu\nu}=0$. For $m=0$, it permits
$(\mathcal{L}_\xi K)^{\mu\nu}=fK^{\mu\nu}$, but contraction with
$(K^{-1})_{\mu\nu}$ gives
$4f=(K^{-1})_{\mu\nu}(\mathcal{L}_\xi K)^{\mu\nu}=-2\partial_\mu\xi^\mu=0$.
Thus both cases give the condition
\begin{equation}
    \mathcal{L}_\xi K^{\mu\nu}=0.
\end{equation}
This is equivalent to the statement that boosts are isometries of the effective metric defining the cone of scalar propagation. This can be defined through the inverse of the kinetic tensor as $H_{\mu\nu}\equiv(K^{-1})_{\mu\nu}$ which reduces to the above result since $K^{\mu\nu}$ is invertible.

In Rindler coordinates, we have that $\xi=\partial_\eta$ with constant coefficients. Consequently, the covariant condition can be expressed as 
\begin{equation}
\mathcal{L}_\xi\mathcal{K}^{mn}=\partial_\eta\mathcal{K}^{mn}=0,
\end{equation}
where $\mathcal{K}^{mn}$ is the kinetic tensor in Rindler coordinates, i.e. $\mathcal{K}^{mn}=K^{\mu\nu}(\partial y^m/\partial x^\mu)(\partial y^n/\partial x^\nu)$. Invariance under boosts is therefore exactly the same requirement that the scalar action and field equation contain no explicit dependence on Rindler time. As long as this condition holds, the field $\varphi$ admits the standard separation into modes of fixed Rindler frequency $\varphi(\eta,\rho,x^A)=\exp{[-i\omega\eta]}\exp{[ip_Ax^A]}\chi_{\omega p_\perp}(\rho)$.

We now solve the covariant boost-invariance condition
$\mathcal{L}_{\xi}K^{\mu\nu}=0$ for the most general constant symmetric
kinetic tensor. In inertial Cartesian coordinates, let
\begin{equation}
K^{\mu\nu}
=
\begin{pmatrix}
A & B & u_{2} & u_{3} \\
B & C & v_{2} & v_{3} \\
u_{2} & v_{2} & D & E \\
u_{3} & v_{3} & E & F
\end{pmatrix}.
\label{eq:general_constant_K}
\end{equation}
The coefficients $A$, $B$, and $C$ describe the components of
$K^{\mu\nu}$ in the $(x^{0},x^{1})$ plane. 
We denote the components
that mix this plane with the transverse directions by
$u_I\equiv K^{0I}$ and $v_I\equiv K^{1I}$, where $I=2,3$. The remaining
components form the symmetric transverse block
$M^{IJ}\equiv K^{IJ}$, with $M^{22}=D$, $M^{23}=M^{32}=E$, and
$M^{33}=F$.

The generator of boosts in the $(x^{0},x^{1})$ plane is
$\xi=x^{1}\partial_{0}+x^{0}\partial_{1}$. Its only nonvanishing
derivatives are $\partial_{1}\xi^{0}=1$ and
$\partial_{0}\xi^{1}=1$. For a contravariant rank-two tensor, the Lie
derivative is
\begin{equation}
\left(\mathcal{L}_{\xi}K\right)^{\mu\nu}
=
\xi^{\alpha}\partial_{\alpha}K^{\mu\nu}
-
K^{\alpha\nu}\partial_{\alpha}\xi^{\mu}
-
K^{\mu\alpha}\partial_{\alpha}\xi^{\nu}.
\label{eq:Lie_derivative_definition}
\end{equation}
Because $K^{\mu\nu}$ is constant in the inertial Cartesian frame, the
first term vanishes. The remaining terms simply exchange the
$x^{0}$ and $x^{1}$ indices. In particular, one obtains
$(\mathcal{L}_{\xi}K)^{00}=-2B$,
$(\mathcal{L}_{\xi}K)^{01}=-(A+C)$,
$(\mathcal{L}_{\xi}K)^{11}=-2B$,
$(\mathcal{L}_{\xi}K)^{0I}=-v_I$,
$(\mathcal{L}_{\xi}K)^{1I}=-u_I$, and
$(\mathcal{L}_{\xi}K)^{IJ}=0$. Collecting these components gives
\begin{equation}
\left(\mathcal{L}_{\xi}K\right)^{\mu\nu}
=
-
\begin{pmatrix}
2B & A+C & v_{2} & v_{3} \\
A+C & 2B & u_{2} & u_{3} \\
v_{2} & u_{2} & 0 & 0 \\
v_{3} & u_{3} & 0 & 0
\end{pmatrix}.
\label{eq:Lie_derivative_matrix}
\end{equation}
The boost-invariance condition therefore requires
$B=0$, $C=-A$, $u_I=0$, and $v_I=0$. Since the time coordinate must
remain timelike with respect to the kinetic tensor, we have
$K^{00}=A<0$. It is consequently convenient to write
$A=-\lambda$, where $\lambda>0$. The most general constant symmetric
kinetic tensor invariant under boosts in the $(x^{0},x^{1})$ plane is
then
\begin{equation}
K^{\mu\nu}
=
\begin{pmatrix}
-\lambda & 0 & 0 & 0 \\
0 & \lambda & 0 & 0 \\
0 & 0 & D & E \\
0 & 0 & E & F
\end{pmatrix}.
\label{eq:general_boost_invariant_K}
\end{equation}
Positive definiteness of the spatial block requires
$\lambda>0$, $D>0$, and $DF-E^{2}>0$. These conditions also imply
$F>0$. The boost symmetry imposes no further restriction on the
transverse block. In particular, the propagation in the transverse
plane may remain anisotropic even though the theory preserves the
boost subgroup acting in the $(x^{0},x^{1})$ plane.

The same result can be verified directly after transforming to
Rindler coordinates,
$x^{0}=\rho\sinh\eta$ and $x^{1}=\rho\cosh\eta$. The transformed
kinetic tensor is defined by
$\mathcal{K}^{mn}
=K^{\mu\nu}(\partial y^{m}/\partial x^{\mu})
(\partial y^{n}/\partial x^{\nu})$, where
$y^{m}=(\eta,\rho,x^{2},x^{3})$. For the tensor in
Eq.~\eqref{eq:general_boost_invariant_K}, its only nonvanishing
components are
\begin{equation}
\mathcal{K}^{\eta\eta}
=
-\frac{\lambda}{\rho^{2}},
\qquad
\mathcal{K}^{\rho\rho}
=
\lambda,
\qquad
\mathcal{K}^{IJ}
=
M^{IJ},
\label{eq:boost_invariant_Rindler_components}
\end{equation}
while
$\mathcal{K}^{\eta\rho}
=\mathcal{K}^{\eta I}
=\mathcal{K}^{\rho I}=0$. None of these components depends on the
Rindler time $\eta$. This is the component realization of the
covariant condition $\mathcal{L}_{\xi}K^{\mu\nu}=0$.
The scalar equation of motion in the right Rindler wedge consequently
takes the stationary form
\begin{equation}
-\frac{\lambda}{\rho^{2}}
\partial_{\eta}^{2}\varphi
+
\frac{\lambda}{\rho}
\partial_{\rho}
\left(
\rho\partial_{\rho}\varphi
\right)
+
M^{IJ}\partial_{I}\partial_{J}\varphi
-
m^{2}\varphi
=
0.
\label{eq:boost_invariant_Rindler_field_equation}
\end{equation}

For the boost-invariant class in Eq.~\eqref{eq:general_boost_invariant_K}, the effective metric takes the form
\begin{equation}
ds_H^2=\frac{1}{\lambda}
\left(-\rho^2d\eta^2+d\rho^2\right)
+(M^{-1})_{IJ}dx^I dx^J,
\end{equation}
The usual Rindler boundaries $x^1=\pm t$ are therefore also horizons for scalar propagation.

Because all coefficients are independent of $\eta$, the field can be
expanded in modes of definite Rindler frequency as
$\varphi(\eta,\rho,x^{I})
=e^{-i\omega\eta}e^{ip_Ix^{I}}
\chi_{\omega,\mathbf{p}_{\perp}}(\rho)$.
Conversely, if any of $B$, $A+C$, $u_I$, or $v_I$ is nonzero, then
$\mathcal{L}_{\xi}K^{\mu\nu}\neq0$. The transformed kinetic tensor
acquires explicit dependence on $\eta$, so Rindler-time translations
are not a symmetry of the field equation. In that case, we can not use the usual
global separation into modes of fixed Rindler frequency.

We note the distinction between boost invariance of the full scalar dynamics from stationarity of the Wightman function along a single detector trajectory. For a pointlike detector accelerated in the $x^1$ direction, we have that $\Delta x^2=\Delta x^3=0$, so a pullback to this trajectory is only sensitive to 
$H_{00}$, $H_{01}$, and $H_{11}$. Since we require the pullback to be stationary (in the boost-invariant case), we require
\begin{equation}
H_{01}=0,
\qquad
H_{11}=-H_{00}.
\label{eq:worldline_stationarity}
\end{equation}
This condition is weaker than boost invariance of the full scalar theory. We discuss the thermal properties of the stationary detector response in the next section.

\section{Finite-time Unruh-DeWitt response}
\label{sec:finitetimeresponse}

Outside of the boost-invariant subset of $K^\mn$ identified above, 
the scalar field does not admit the standard global decomposition into definite-frequency modes. The response of an accelerated Unruh-DeWitt detector nevertheless remains well-defined since it can be computed directly from the inertial frame Wightman function in found in Section~\ref{sec:scalarfield_quantisation}. In this section we place the Lorentz violation in the scalar-field sector and take the detector itself to follow the standard Rindler trajectory with uniform acceleration.

\subsection{Generic detector response}

For a detector with energy gap $\Omega$ endowed with a real switching function $\chi(\tau)$, 
we can write the leading-order detector response function as~\cite{dewitt1979quantum,UnruhWald84,Dickinson_2025}
\begin{equation}
    \mathcal{F}_\chi(\Omega)=|\mathfrak{m}|^2 \int d\tau d\tau^\prime \chi(\tau)\chi(\tau^\prime)e^{-i\Omega(\tau-\tau^\prime)}W^+(\tau,\tau^\prime).
    \label{eq:switched_response}
\end{equation}
Here $|\mathfrak{m}|^2 = |\langle E_2| {\hat M} (0) | E_1 \rangle|^2$, 
referring to the operator $\hat M$ in \rf{Sint}, and $\Om=E_2-E_1$.
This expression is valid regardless of whether the pulled-back Wightman function is stationary. 
In the nonstationary case, $\mathcal F_\chi(\Omega)$ is the fundamental detector observable and cannot generally be written as the observation time multiplied by a time-independent transition rate as in the standard case. For convenience, we use the effective metric $H_{\mu\nu}$ defined as by $H_{\mu\alpha}K^{\alpha\nu}=\delta_\mu^{~\nu}$
so that the coordinate-space two-point functions are the ordinary Minkowski ones but evaluated on the quadratic form given by $H_{\mu\nu}$. 
We use the accelerated worldline in the $x^1$ direction (see Eq.~\eqref{wl0}). 
We can then define the invariant interval pulled back to the detector worldline
\begin{equation}
\label{eq:QK_boost_breaking}
\begin{aligned}
    Q_{K,\epsilon}(\Sigma,\Delta)=-\frac{4\mathcal{X}(\Sigma)}{a^2}\sinh^2{\frac{a\Delta}{2}}+\frac{4i\epsilon}{a\sqrt{\ga}}\cosh{a\Sigma}\sinh{\frac{a\Delta}{2}}+\epsilon^2,
\end{aligned}
\end{equation}
which is derived in Appendix~\ref{app:pullbackiepsilon}. Here, we have used $\Sigma\equiv (\tau+\tau^\prime)/2$, $\Delta\equiv\tau-\tau^\prime$ and
\begin{equation}
    \mathcal{X}(\Sigma)=-\frac{1}{2}(H_{00}+H_{11})\cosh{(2a\Sigma)}-H_{01}\sinh{(2a\Sigma)}-\frac{1}{2}(H_{00}-H_{11}).
    \label{chH}
\end{equation}
The condition $\mathcal X(\Sigma)>0$ is equivalent to $H_{\mu\nu}\dot x^\mu\dot x^\nu<0$. It ensures that the detector trajectory is timelike with respect to scalar propagation. The observable is the finite-time transition probability $\mathcal{F}_{\chi}(\Omega)$ for specified detector settings. As discussed in Section~\ref{eft}, coordinate changes can move coefficients between the scalar and detector sectors, but leave this probability unchanged

The pullback is stationary when $H_{01}=0$ and $H_{11}=-H_{00}$ and is the same result as Eq.~\eqref{eq:worldline_stationarity} but obtained using the detector method. 
If we do not impose stationarity, the Wightman function generally depends on both $\Sigma$ and $\Delta$. The detector therefore has a nonstationary response, and the usual equilibrium KMS interpretation does not apply, but the finite-time response \eqref{eq:switched_response} remains well defined without assuming a thermal state.
It is interesting to note that if $H_{01}\neq0$ in Eq.~\eqref{eq:QK_boost_breaking}, then $\sinh{(2a\Sigma)}$ remains in the response. This term is odd under $\Sigma\to-\Sigma$ and around the turning point $\Sigma=0$.

We start from the switched detector response in Eq.~\eqref{eq:switched_response}.
Since the Jacobian for the change of variables $\tau \to \Sigma, \Delta$ is unity, 
we can write $d\tau d\tau^\prime=d\Sigma d\Delta$ and
\begin{equation}
    \mathcal{F}_\chi(\Omega)=|\mathfrak{m}|^2 \int d\Sigma d\Delta \chi\left(\Sigma+\frac{\Delta}{2}\right)\chi\left(\Sigma-\frac{\Delta}{2}\right)e^{-i\Omega\Delta}W^+(\Sigma, \Delta).
    \label{eq:switched_response2}
\end{equation}
In the stationary limit, the pulled-back Wightman function is independent of $\Sigma$ and the integration over $\Si$
can be separated from the relative time dependence $\De$. 
A time-independent transition rate can then be obtained in the appropriate long interaction limit. Outside of the stationary sector, this factorization is not available; nevertheless, the finite-time response $\mathcal{F}_\chi(\Omega)$ is still defined as above and generally depends on both the duration and temporal location of the measurement. In order to retain information about when the measurement is performed, we write the switching function as
\begin{equation}
\chi_{T,\tau_c}(\tau)=
\chi_T(\tau-\tau_c),
\quad
\chi_T(-u)=\chi_T(u),
\label{eq:centered_switching}
\end{equation}
where $T$ characterizes the duration of the measurement and $\tau_c$ specifies the center of the measurement window relative to the 
inertial frame. 
Then, the switching functions in Eq.~\eqref{eq:centered_switching} can be written as $\chi_T(\Sigma-\tau_c+\Delta/2)\chi_T(\Sigma-\tau_c-\Delta/2)$, showing exactly that the window is centered around $\Sigma\simeq\tau_c$ and that the Lorentz-violating factor depends on the average proper time $\Sigma$. 
In the stationary limit, $\mathcal{X}=\text{const.}$ and $\Sigma\to\Sigma-\tau_c$ eliminates all dependence on the measurement time $\tau_c$.

For convenience, 
we factor out the invariant interval
for the standard case and write the invariant as
\begin{equation}\label{eq:QK_Lambda}
     Q_K(\Sigma,\Delta)=\mathcal{X}(\Sigma)Q_{\rm std}(\Delta), \quad Q_{\rm std}(\Delta)\equiv-\frac{4}{a^2}\sinh^2{\left[\frac{a}{2}(\Delta-i0)\right]}, 
 \end{equation}
 where $Q_{\rm std}$ represents the Lorentz-invariant case. This factorized form refers to the $\epsilon\to0^+$ limit 
 and is valid for $\mathcal{X}>0$ over the support of the detector switching function. 
 In the $\mathcal{X}(\Sigma)\to\text{ const.}$ limit, we obtain stationarity of the pullback along the accelerator trajectory.\\[2mm]

\noindent\underline{\textit{Massless field}}:\\

In the massless case, the positive Wightman function for general $K^{\mu\nu}$ is
\begin{equation}
    W_0^+(\Sigma,\Delta)=\frac{1}{\sqrt{-\det K}\mathcal{X}(\Sigma)}W_{0,\rm{std}}^+(\Delta;a), 
\end{equation}
where
\begin{equation}
    W_{0,\rm{std}}^+(\Delta;a)\equiv-\frac{a^2}{16\pi^2}\frac{1}{\sinh^2{\left[\frac{a}{2}(\Delta-i0)\right]}},
\end{equation}
is the ordinary massless Wightman function. Substituting the pulled-back Wightman function into the switched response \eqref{eq:switched_response} gives
\begin{align}
\mathcal F_{\chi,0}(\Omega)=|\mathfrak{m}|^2\frac{1}{\sqrt{-\det K}}\int d\Sigma\,d\Delta\,
\chi\left(\Sigma+\frac{\Delta}{2}\right)\chi\left(\Sigma-\frac{\Delta}{2}\right)
\frac{e^{-i\Omega\Delta}}{\mathcal X(\Sigma)}
W_{0,\mathrm{std}}^+(\Delta;a).
\label{eq:massless_finite_response}
\end{align}\\[2mm]

\noindent\underline{\textit{Massive field}}:\\

When the field is massive, we use the massive Wightman function in Minkowski space given in Eq.~\eqref{twopointm3}. Assuming that $\mathcal{X}(\Sigma)>0$, we can factorise the invariant interval $Q_K$ as in the massless case. Then, we can write the  pulled back Wightman function as
\begin{equation}\label{eq:wightmanplus}
    W_m^+(\Sigma,\Delta)=\frac{1}{\sqrt{-\det K}\mathcal{X}(\Sigma)}\left[\frac{m\sqrt{\mathcal{X}(\Sigma)}K_1\left(m\sqrt{\mathcal{X}(\Sigma)}\sqrt{Q_{\rm std}(\Delta)}\right)}{4\pi^2\sqrt{Q_{\rm std}(\Delta)}}\right],
\end{equation}
where the square bracket is exactly the ordinary Wightman function in the massive case, but with an effective time-dependent mass as $m \to \mu(\Sigma)=m\sqrt{\mathcal{X}(\Sigma)}$. We can therefore write the final result as
\begin{equation}
    W_m^+(\Sigma,\Delta)=\frac{1}{\sqrt{-\det K}\mathcal{X}(\Sigma)}W^{\rm std}_{m\sqrt{\mathcal{X}(\Sigma)}}(\Delta;a),
\end{equation}
using which we can write the exact switched massive response as
\begin{align}
\mathcal F_{\chi,m}(\Omega)=|\mathfrak{m}|^2\frac{1}{\sqrt{-\det K}}\int d\Sigma\,d\Delta\,
\chi\left(\Sigma+\frac{\Delta}{2}\right)
\chi\left(\Sigma-\frac{\Delta}{2}\right)
\frac{e^{-i\Omega\Delta}}{\mathcal X(\Sigma)}
W^{\mathrm{std},+}_{m\sqrt{\mathcal X(\Sigma)}}(\Delta;a).
\label{eq:massive_finite_response}
\end{align}
For the boost-invariant class
\eqref{eq:general_boost_invariant_K}, we have
$\mathcal X=1/\lambda$ and
$\det K=-\lambda^2(DF-E^2)$. The pulled-back Wightman
function therefore becomes
$W_m^+(\Delta)=W_{m/\sqrt{\lambda}}^{\rm std,+}(\Delta;a)/\sqrt{DF-E^2}$.
Since the standard accelerated Wightman function satisfies
the KMS condition with inverse temperature $\beta=2\pi/a$,
the same holds here for both massive and massless fields.
More generally, the weaker condition
\eqref{eq:worldline_stationarity} also makes
$\mathcal X$ constant and gives a KMS two-point function
along the detector trajectory, without requiring full
boost invariance or the usual scalar-field Rindler horizons.

\subsection{
Weak boost breaking and measurement-time dependence}
\label{weak boost breaking}

As a preliminary investigation of the
spectrum, 
we consider the case when the boost invariance is only weakly broken, we define $\mathcal{X}_0\equiv\mathcal{X}(0)=-H_{00}$ and we take
 \begin{equation}
 \mathcal{X}(\Sigma)=\mathcal{X}_0+\delta\mathcal{X}(\Sigma), \quad \mathcal{X}_0=-H_{00}>0, \quad \left|\frac{\delta\mathcal{X}_0}{\mathcal{X}_0}\right|\ll1,
 \end{equation}
so from Eq.\ \eqref{eq:QK_Lambda} we identify
 \begin{equation}\label{eq:Xexpand}
     \delta\mathcal{X}(\Sigma)=-\frac{1}{2}(H_{00}+H_{11})[\cosh{(2a\Sigma)}-1]-H_{01}\sinh{(2a\Sigma)}.
 \end{equation}
 For the massive case, 
 we can define the dressed mass to zeroth order as $\mu_0\equiv m\sqrt{\mathcal{X}_0}$, 
 after which we can write the detector response as
 \begin{equation}
     \frac{1}{\mathcal{X}(\Sigma)}W_{m\sqrt{\mathcal{X}(\Sigma)}}^{\rm{std},+}(\Delta;a)=\frac{1}{\mathcal{X}_0}\Bigg[W_{\mu_0}^{\rm{std,+}}(\Delta;a)+\frac{\delta\mathcal{X}(\Sigma)}{\mathcal{X}_0}\left(\frac{\mu_0}{2}\frac{\partial}{\partial\mu} W_{\mu}^{\rm{std,+}}(\Delta;a)\Big|_{\mu=\mu_0}-W_{\mu_0}^{\rm{std,+}}(\Delta;a)\right)\Bigg]+\mathcal{O}(\delta\mathcal{X}^2),
 \end{equation}
 where we see that in the boost-preserving limit, we have $H_{00}=-H_{11}$ and $H_{01}=0$, consistent with the previous section. For a massless field, the boost breaking modifies the response through the factor $1/\mathcal X(\Sigma)$; furthermore, for a massive field, there is an additional correction from the $\Sigma$-dependent effective mass. With a finite switching function, neither detector response can in general be identified with an exactly Planckian spectrum
\footnote{For a massless field with Gaussian switching,
the Lorentz-violating response differs from the
conventional finite-time response only by an overall
factor. The finite-time spectrum is not, however,
exactly Planckian.}.

Assuming now a measurement centered around $\tau_c$, we define $u\equiv\Sigma-\tau_c$, and the switching product as $\chi_T(u+\Delta/2)\chi_T(u-\Delta/2)$, which is even under $u\to-u$. Then, we take the odd part of $\delta\mathcal{X}$ defined in Eq.~\eqref{eq:Xexpand} and rewrite it using trigonometric identities as
\begin{equation}
    \sinh{[2a(\tau_c+u)]}=\sinh{(2a\tau_c)}\cosh{(2au)}+\cosh{(2a\tau_c)}\sinh{(2au)}.
\end{equation}
When integrated over $u$, the second term vanishes, and the surviving contribution is $\sinh{(2a\tau_c)}$. Hence, we can write the part of the massive response function that is antisymmetric under $\tau_c\to-\tau_c$ as
\begin{equation}
    \begin{aligned}
        \frac{1}{2}(\mathcal{F}_{\chi,m}(\Omega;\tau_c)-\mathcal{F}_{\chi,m}(\Omega;-\tau_c))=|\mathfrak{m}|^2 \frac{H_{01}\sinh{(2a\tau_c)}}{\sqrt{-\det K}\mathcal{X}_0^2}&\int du\, d\Delta\cosh{(2au)}\chi\left(u+\frac{\Delta}{2}\right)\chi\left(u-\frac{\Delta}{2}\right)e^{-i\Omega\Delta}\\&\;\hspace{-2em}\times\left[W_{\mu_0}^{\rm{std,+}}(\Delta;a)-\frac{\mu_0}{2}\frac{\prt}{\prt \mu} W_{\mu}^{\rm{std,+}}(\Delta;a)\Big|_{\mu=\mu_0}\right]+\mathcal{O}(\delta\mathcal{X}^2),
    \end{aligned}
\end{equation}
where we see that a measurement centered around $\tau_c=0$ loses the first-order correction from $H_{01}$ and that $H_{01}$ contributes with the opposite sign for $\tau_c$ and $-\tau_c$, and that $H_{00}+H_{11}$ contributes to the even part of the measurement. 

\begin{figure}[t]
\begin{tikzpicture}[
    x=1cm,y=1cm,
    >=Latex,
    line cap=round,
    line join=round
]

\path[use as bounding box] (0,0) rectangle (16,8);

\draw[dashed,semithick]
  (1.05,4.95) .. controls (3.4,5.75) and (11.8,5.55) .. (15.0,6.35);

\foreach \yy in {4.10,4.30,4.50}{
  \draw[black!25,thin]
    plot[smooth,domain=0.35:15.75,samples=180]
    (\x,{\yy + 0.10*sin(120*\x)});
}

\foreach \x in {0.45,0.70,...,15.55}{
  \foreach \y in {3.70,3.95,4.20,4.45,4.70}{
    \fill[black!60] (\x,\y) circle (0.010);
  }
}

\newcommand{\rocket}[6][0.72]{%
  \begin{scope}[
    shift={(#2,#3)},
    scale=#1,
    rotate=#4,
    line cap=round,
    line join=round
  ]

    \draw[
      fill=white,
      line width=0.65pt
    ]
      (-0.69, 0.16)
      .. controls (-0.94, 0.31)
                and (-1.17, 0.31)
             .. (-1.39, 0.14)
      -- (-1.22, 0.04)
      -- (-1.56,-0.13)
      -- (-1.18,-0.15)
      -- (-1.40,-0.37)
      .. controls (-1.10,-0.36)
                and (-0.87,-0.27)
             .. (-0.66,-0.15)
      .. controls (-0.58,-0.04)
                and (-0.59, 0.07)
             .. (-0.69, 0.16)
      -- cycle;

    \draw[
      fill=white,
      line width=0.65pt
    ]
      (-0.46,0.32)
      -- (-0.74,0.52)
      -- (-0.99,0.40)
      -- (-0.68,0.17)
      -- cycle;

    \draw[
      fill=white,
      line width=0.65pt
    ]
      (-0.30,-0.38)
      -- (-0.52,-0.68)
      -- (-0.86,-0.70)
      -- (-0.66,-0.27)
      -- cycle;

    \draw[
      fill=white,
      line width=0.70pt
    ]
      (-0.70,0.18)
      .. controls (-0.37,0.44)
                and ( 0.26,0.58)
             .. ( 0.96,0.50)
      .. controls (1.28,0.46)
                and (1.52,0.29)
             .. (1.68,0.08)
      .. controls (1.53,-0.15)
                and (1.28,-0.34)
             .. (0.95,-0.43)
      .. controls ( 0.23,-0.63)
                and (-0.38,-0.53)
             .. (-0.60,-0.34)
      .. controls (-0.69,-0.24)
                and (-0.76,0.04)
             .. (-0.70,0.18)
      -- cycle;

    \draw[line width=0.70pt]
      (1.02,0.49)
      .. controls (0.98,0.23)
                and (1.02,-0.08)
             .. (1.14,-0.37);

    \begin{scope}[shift={(0.34,0.02)}]
      \draw[
        fill=white,
        line width=0.70pt
      ]
        (0,0) circle (0.33);

      \foreach \ang in {0,30,...,330}{
        \draw[line width=0.45pt]
          (\ang:0.265) -- (\ang:0.31);
      }

      \draw[line width=0.75pt]
        (0,0) -- (#5:0.225);

      \draw[line width=0.75pt]
        (0,0) -- (#6:0.17);

      \fill (0,0) circle (0.022);
    \end{scope}

  \end{scope}%
}

\rocket[0.72]{4.0}{5.55}{9}{100}{155}
\rocket[0.72]{8.2}{5.70}{8}{35}{110}
\rocket[0.72]{12.5}{6.10}{9}{300}{20}

\node[align=center] at (4.0,6.45)
  {$\tau_c<0$};

\node[align=center] at (8.2,6.48)
  {$\tau_c=0$};

\node[align=center] at (12.5,6.82)
  {$\tau_c>0$};

\draw[dashed,semithick] (4.0,5.15) -- (4.0,3.15);
\draw[dashed,semithick] (8.2,5.25) -- (8.2,3.15);
\draw[dashed,semithick] (12.5,5.60) -- (12.5,3.15);

\newcommand{\panel}[5]{%
  \begin{scope}[shift={(#1,#2)}]


    \draw[->,semithick] (-1.35,-0.55) -- (-1.35,0.78);
    \draw[->,semithick] (-1.35,-0.55) -- (1.45,-0.55);

    \node[anchor=south west] at (-1.40,0.82)
      {$\Gamma/(a|\mathfrak{m}|^2)$};

    \node[anchor=north] at (-1.35,-0.56)
      {$0$};

    \node[anchor=west] at (1.48,-0.55)
      {$\Omega/a$};

    \draw[
      black,
      densely dashed,
      semithick
    ]
      plot[smooth] coordinates {#4};

    \draw[
      black,
      very thick
    ]
      plot[smooth] coordinates {#5};

    \node at (0,-1.00) {#3};

  \end{scope}
}

\panel{4.0}{1.75}{suppressed}
{
  (-1.28,-0.52)
  (-1.00, 0.12)
  (-0.72, 0.40)
  (-0.35, 0.42)
  ( 0.05, 0.20)
  ( 0.50,-0.05)
  ( 0.95,-0.28)
  ( 1.25,-0.40)
}
{
  (-1.28,-0.52)
  (-1.02,-0.02)
  (-0.75, 0.15)
  (-0.40, 0.10)
  ( 0.00,-0.10)
  ( 0.45,-0.28)
  ( 0.90,-0.40)
  ( 1.25,-0.48)
}

\panel{8.2}{1.75}{thermal-like}
{
  (-1.28,-0.52)
  (-1.00, 0.08)
  (-0.72, 0.34)
  (-0.35, 0.38)
  ( 0.02, 0.18)
  ( 0.45,-0.06)
  ( 0.88,-0.28)
  ( 1.22,-0.42)
}
{
  (-1.28,-0.52)
  (-1.02, 0.03)
  (-0.72, 0.31)
  (-0.35, 0.35)
  ( 0.03, 0.16)
  ( 0.45,-0.03)
  ( 0.88,-0.27)
  ( 1.22,-0.41)
}

\panel{12.5}{1.75}{enhanced}
{
  (-1.28,-0.52)
  (-1.00, 0.10)
  (-0.72, 0.30)
  (-0.35, 0.28)
  ( 0.02, 0.08)
  ( 0.45,-0.18)
  ( 0.88,-0.36)
  ( 1.22,-0.46)
}
{
  (-1.28,-0.52)
  (-1.03, 0.18)
  (-0.75, 0.55)
  (-0.38, 0.58)
  ( 0.02, 0.38)
  ( 0.48, 0.05)
  ( 0.92,-0.18)
  ( 1.22,-0.28)
}
\end{tikzpicture}
\caption{Schematic picture of measurement time dependence of the Unruh effect due to nonzero coefficients for Lorentz violation.}
\label{fig:rocket}
\end{figure}
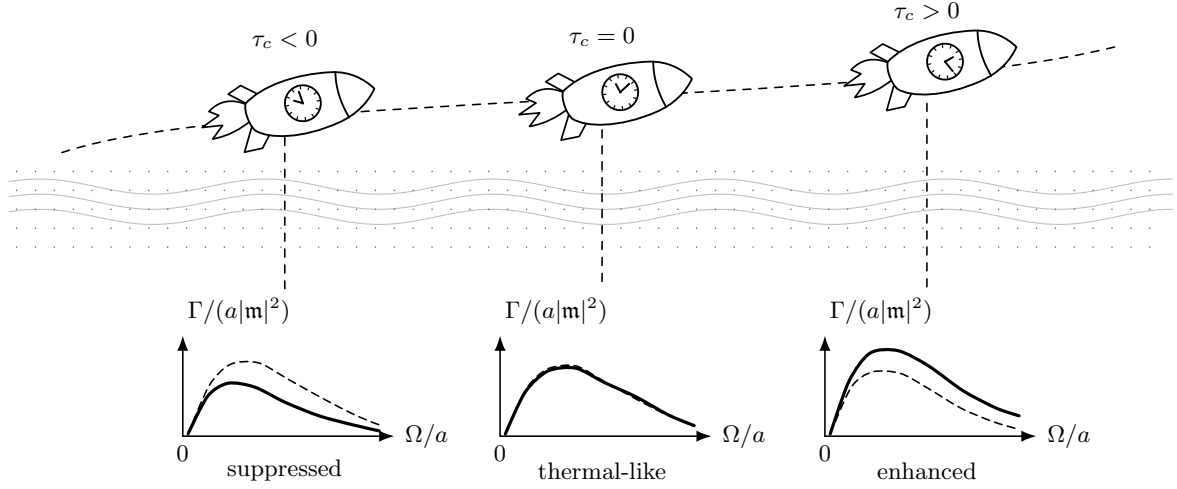

In order to estimate this effect analytically, we temporarily choose a Gaussian switching function of the form
\begin{equation}
\chi_{T,\tau_c}(\tau)=\exp\left[-\frac{(\tau-\tau_c)^2}{2T^2}\right].
\label{eq:Gaussian_switching}
\end{equation}
Then, the normalized detector response (or excitation rate), 
defined as $\overline\Ga=\mathcal{F}_{\chi,m}(\Omega;\tau_c)/(\sqrt{\pi}T)$, 
reads
\begin{equation}
\overline\Gamma_{m,T}(\Omega;\tau_c)=\frac1{\sqrt{\pi}T\sqrt{-\det K}}\int_{-\infty}^{\infty}d\Sigma\,\frac{e^{-(\Sigma-\tau_c)^2/T^2}}{\mathcal{X}(\Sigma)}
\Gamma_{m\sqrt{\mathcal{X}(\Sigma)},T}^{\rm std}(\Omega;a),
\label{eq:exact_Gaussian_response}
\end{equation}
where $\Gamma_{\mu,T}^{\rm std}(\Omega;a)$ is the standard Lorentz-invariant detector response for the Gaussian switching, defined as
\begin{equation}
\Gamma_{\mu,T}^{\rm std}(\Omega;a)\equiv |\mathfrak{m}|^2\int_{-\infty}^{\infty}d\Delta\,e^{-\Delta^2/(4T^2)}e^{-i\Omega\Delta}W_{\mu}^{\rm std,+}(\Delta;a).
\label{eq:standard_Gaussian_response}
\end{equation}
Then, to linear order in $\delta\mathcal{X}$, we can write 
\begin{equation}
    \frac{1}{\mathcal{X}(\Sigma)}\Gamma_{m\sqrt{\mathcal{X}},T}^{\rm std}=\frac{1}{\mathcal{X}_0(\Sigma)}\left[\Gamma_{\mu_0,T}^{\rm std}+\frac{\delta\mathcal{X}(\Sigma)}{\mathcal{X}_0}\left(\frac{\mu_0}{2}
    \frac{\prt}{\prt \mu} \Gamma_{\mu,T}^{\rm std}\Big|_{\mu=\mu_0}-\Gamma_{\mu_0,T}^{\rm std}\right)\right].
\end{equation}
We compute averages over the measurement period by applying a Gaussian weighted average defined as
\begin{equation}
    \langle f\rangle_T\equiv \frac{1}{\sqrt{\pi}T}\int_{-\infty}^\infty d\Sigma \;e^{-(\Sigma-\tau_c)^2/T^2}f(\Sigma),
\end{equation}
using which we see that
\begin{equation}
\langle\cosh{(2a\Sigma)}\rangle_T=e^{a^2T^2}\cosh{(2a\tau_c)},\quad \langle\sinh{(2a\Sigma)}\rangle_T=e^{a^2T^2}\sinh{(2a\tau_c)},
\end{equation}
and we can write the average of the weak boost-breaking function \eqref{eq:Xexpand} as 
\begin{equation}
\langle\delta\mathcal{X}\rangle=-\frac{1}{2}(H_{00}+H_{11})\left[e^{a^2T^2}\cosh{(2a\tau_c)-1}\right]-H_{01}e^{a^2T^2}\sinh{(2a\tau_c)},
\end{equation}
Using, this, we can write the
we can write the massive response function to first order as
 \begin{align}
\overline\Gamma_{m,T}(\Omega;\tau_c)
=
\frac1{\sqrt{-\det K}\,\mathcal{X}_0}
\Bigg\{
\Gamma_{\mu_0,T}^{\rm std}+&
\frac{H_{00}+H_{11}}{2\mathcal{X}_0}
\left[
e^{a^2T^2}\cosh(2a\tau_c)-1
\right]
\left[
\Gamma_{\mu_0,T}^{\rm std}
-
\frac{\mu_0}{2}
\frac{\prt}{\prt \mu}
\Gamma_{\mu,T}^{\rm std}
\bigg|_{\mu=\mu_0}
\right]
\nonumber\\
&+
\frac{H_{01}}{\mathcal{X}_0}
e^{a^2T^2}\sinh(2a\tau_c)
\left[
\Gamma_{\mu_0,T}^{\rm std}
-
\frac{\mu_0}{2}
\frac{\partial}{\partial\mu}\Gamma_{\mu,T}^{\rm std}
\bigg|_{\mu=\mu_0}
\right]
\Bigg\}
+
O(\delta\mathcal{X}^2).
\label{eq:massive_small_breaking_response}
\end{align}
Because $\delta\mathcal X(\Sigma)$ contains exponentially growing hyperbolic functions, small tensor coefficients do not by themselves guarantee perturbative validity for long measurements. We must also impose that $\left|\delta\mathcal X(\Sigma)/\mathcal X_0\right|\ll1$ throughout the whole region where the switching function has support. 
This calculation describes the general weakly boost breaking response and includes both the even and odd contributions, controlled by $H_{00}+H_{11}$ and $H_{01}$ respectively. A schematic illustration of the measurement time dependence can be seen in Figure~\ref{fig:rocket}.

\section{Numerical solutions of the finite-time response}
\label{detector}

For what follows, 
we use the standard trajectory in Eq.\ \rf{wl0} and we use Eq.~\eqref{eq:switched_response} for the finite-time detector response.
In this section, 
we recast the response function $\cal F$ 
in a form suitable for numerical evaluations of the response function and excitation rate.

\subsection{Integral set-up}
\label{integral set-up}

In Eq.~\eqref{eq:switched_response}, the Wightman function is to be evaluated along two Rindler trajectories with different proper times $\tau$ and $\tau^\prime$. We can therefore express the Wightman function as $W=W[x(\tau),y(\tau^\prime)]$, where $x$ is defined above in \eqref{wl0}.
For $y^\mu$ we use the same trajectory but evaluated at $\ta^\prime$. Using hyperbolic identities, 
we can write the components of $z^\mu=x^\mu-y^\mu$ as in Eq.~\eqref{eq:coordinate_separation_accelerated_worldline}.
It is known that a more general (regulator) shift can be used to simplify the expression (e.g., see Ref.\ \cite{Alves:2023jmc} and references therein). 
The idea is to employ the shift (via analytic continuation)
\begin{equation}
z^\mu \rightarrow z^\mu - i \ep T^\mu,
\end{equation}
where the components of the vector $T^\mu$ are left unspecified
except that they are independent of the momentum $\tilde p$
in Eq.\ \rf{twopointm2}.
However, 
$T^\mu$ is also required to be timelike, 
or more speficially, 
we must have $T^0 > |\bf T|$, 
in order for the integral \rf{twopointm2} to converge.
We therefore use a revised $Q$ given by
\beq
Q_\ep = ( {\mathbf {\tilde z}} - i \ep {\mathbf T})^2 - ({\tilde z^0} - i \ep T^0)^2
\label{Qreplace}.
\eeq

Since the general massless Wightman function appears to be essentially a time-dependent scaling of the standard result, 
we focus on the massive field. 
For this case,
upon using the generic $i\ep T^\mu$ shift in Eq.~\rf{Qreplace},
the transition probability becomes,
\beq
\mathcal{F}_\chi(\Omega)=
|\mathfrak{m}|^2
\int_{-\infty}^{\infty} \int_{-\infty}^{\infty} 
\ch (\ta) \ch(\ta^\prime) 
\frac{m}{4\pi^2 \sqrt{-det K} } 
\frac {K_1 (m \sqrt{Q_\epsilon})}{\sqrt{Q_\epsilon}}
e^{-i \Om \De} d \ta d \ta^\prime.
\label{amp2}
\eeq
The trajectory \rf{wl0} is inserted into $Q_\ep$
and a choice for $T^\mu$ is made.
Specifically, 
we let the components of $T^\mu$ be given by,
\beq
T^0 = \frac {1}{a \sqrt{\ga} } \cosh a\ta,
\pt{spa} 
T^j = \frac {1}{a} \left(M^j_{\pt{j}1} \sinh a\ta + M^j_{\pt{j}l} \frac{K^{0l}}{\ga} \cosh a\ta \right).
\label{Tchoice}
\eeq
After simplication, 
$Q$ becomes
\beq
Q_{\ep,T} = -\frac{4}{a^2} \sinh^2 (a \frac{\De}{2}-i\ep)
\left[  
 -\frac{1}{2}(H_{00}+H_{11})\cosh{(2a\Sigma-i\ep)}-H_{01}\sinh{(2a\Si -i \ep)}-\frac{1}{2}(H_{00}-H_{11}).
\right].
\label{Qr}
\eeq
Note that we have used the relations between $M^i_{\pt{i}j}$
and $H_\mn$ given in Appendix \ref{relations}.

Collecting these equations and simplifying Eq.~\rf{amp2} we obtain
\beq
\mathcal{F}_\chi(\Omega)=
\frac{-i m a |\mathfrak{m}|^2} {8\pi^2 \sqrt{-\det K \ze}}  
\int_{-\infty}^{\infty} 
\int_{-\infty}^{\infty} 
\ch (\ta) \ch(\ta^\prime) 
\frac {1}{\cR}
K_1 \left( 
i \frac {2m \sqrt{\ze}}{a}  \cR 
\right)
e^{-i \Om \De} d \ta d \ta^\prime, 
\label{amp3}
\eeq
where 
\beq
{\cal R} = \sinh \left( \frac{a\De}{2}-i\ep \right) 
\sqrt{1+\de \cosh (2 a \Si -i\ep) + \ka \sinh (2 a \Si -i\ep) }.
\label{R}
\eeq
Note that in simplifying \rf{Qr}, 
we have made a choice of branch for the square root of the complex expression $Q_{\ep,T}$.
In addition to the $\det K$, the quantities $\ze$, $\de$, and $\ka$ contain all the of the coefficient dependence.
Specifically, 
these quantities are given by,
\beq
\bal
\ze &= -\frac 12 (H_{00}-H_{11}) \approx 1+\frac 12 (k_{00}-k_{11}), 
\\
\de &= \frac {H_{00}+H_{11}}{H_{00}-H_{11}} \approx \frac 12 (k_{00}+k_{11}),
\\
\ka &=\frac {2 H_{01}}{H_{00}-H_{11}} \approx k_{01},
\label{zedeka}
\eal
\eeq
where the leading order expansions in terms of $k_\mn$ coefficients are shown for 
clarity (but not assumed for what follows).
Note also that $-\det K \approx 1+k^\mu_{\pt{\mu}\mu}$.
In order that the argument of the square root term in $\cal R$ remain positive it suffices that $\de>|\ka|$.
The modifications of the $\cal F$ 
due to the symmetry-breaking action \rf{scalar_act}, 
evidently depend 
on three coefficients at leading order.

One key observable is the transition rate, 
or how many excitations occur per unit time (from the ground state to the excited state for a two-level quantum system).
One common approach that suffices for the conventional case with $k_\mn=0$, 
is to take $\ch (\ta)$ as a step function with finite duration
\cite{Crispino2008,Alves:2023jmc}.
In the present case, 
we have a nonstationary integrand that depends on both $\De=\ta-\ta^\prime$ and $\Si=(\ta+\ta^\prime)/2$.
When one changes variables in the integral from $(\ta, \ta^\prime)$ to $(\De,\Si)$, 
as in Eq.\ \rf{eq:switched_response2}, 
step functions $\chi$ and $\chi^\prime$ transform
the integration region from a ``top hat" to a diamond shape.
This changes the integral to, 
schematically,
\beq
{\cal F} \propto \int_{-\De_m}^{\De_m} 
\int_{-\De_m+|\De|}^{\De_m-|\De|} f(\De, \Si) d\De d \Si.
\label{tophatfail}
\eeq
Since the integrand has a (regulated) pole at $\De=0$, 
the non-differentiable behavior of $|\De|$ is problematic, 
as we have discovered with numerical integration.  
Only when the integrand is $\Si$-independent, 
is the conventional result recovered without singularity issues.

Instead of the step function approach, 
we use Gaussian switching functions.
We calculate the amplitude for the transition and then divide by a measure of the time interval captured by the gaussian 
to get a result we use
as the excitation rate.
This is similar to the ``tophat" approach in the conventional case,
where the rate $\Ga = {\cal F_\ch}/{\cal T}$, 
with $\cal T$ being the duration for the detector.
Specifically, 
we take $\ch$ to be
\beq
\ch (\ta) = e^{-(\ta-\ta_c)^2/2\si^2}
\label{gauss_chi},
\eeq
where $\ta_c$ is the central time for the detection as discussed in Section \ref{weak boost breaking}, 
and $\si$ is the width of the distribution.
We next return to \rf{amp3}, 
and insert \rf{gauss_chi}. 
We also change variables from $\ta,\tap$ to the dimensionless
$u=a \De/2$ and $v=a\Si$, 
yielding
$d\ta d\tap = 2du dv /a^2$, 
while the region of integration remains the infinite domain.
With these changes, 
the amplitude becomes
\beq
\mathcal{F}_g(\Omega)=
\frac{-i|\mathfrak{m}|^2} {4\pi^2 \sqrt{-\det K \ze}}
\frac{m}{a}
\int_{-\infty}^{\infty}
\int_{-\infty}^{\infty}
du dv \;
{\rm Exp}\left[ -\frac {1}{(a \si)^2} (u^2 + (v-v_c)^2)\right] 
\frac {1}{\cR}
K_1 \left( 
i \frac {2m \sqrt{\ze}}{a}  \cR 
\right)
e^{-i 2 \Om u/a}, 
\label{amp4}
\eeq
where $v_c=a\ta_c$, 
and the $g$ subscript indicates a Gaussian switching function.
To obtain a measure of the excitation rate $\Ga$, 
we use
\beq
\Ga_{\rm avg} = \frac {{\cal F}_g}{\sqrt{\pi}\si}.
\label{rate}
\eeq

It is worth distinguishing the $a\to0$ and $m/a\gg1$ limits of the Gaussian response in Eq.~\eqref{amp4}. In the first case and with fixed $\{m, \Omega,\sigma, \tau_c\}$, the apparent singularity disappears when we return to the original proper-time variables; indeed, using the definition of $\mathcal{R}$ in Eq.~\eqref{R}, we find that all explicit powers of $a$ cancel. The resulting Wightman function depends only on $\tau-\tau^\prime$ so the response becomes independent of $\tau_c$; However, Lorentz violation remains through the quantities $\zeta(1+\delta)$ and det $K$ as can be seen from Eqs.~\eqref{chH} and \eqref{zedeka}. The $a\to0$ limit does therefore not reproduce the Lorentz-invariant inertial response in general. Furthermore, the Gaussian switching function in Eq.~\eqref{eq:massive_finite_response} give a nonzero excitation probability even at zero acceleration. The $m/a\gg1$ limit, which is to be understood as taken for fixed $a$, shows different behaviour. In the massive response function in Eq.~\eqref{eq:massive_finite_response}, we find exponential suppression in this limit with the mass dependence going as $\mathcal{F}_g\propto\text{exp}[-2m\sqrt{\zeta}/a]$ to leading order, at least when the pullback is stationary. Here, Lorentz violation again remains through $\zeta$ and det $K$. In the nonstationary case, the effective mass $m\sqrt{\mathcal{X}(\Sigma)}$ depends on the averaged proper time and the resulting suppression can also depend on $\tau_c$. To isolate the effects which are strictly from acceleration, one should in principle subtract the inertial ($a\to0)$ response from the finite $a$ result as discussed e.g. in Ref.~\cite{Dickinson_2025}. We leave this subtraction for future work and instead use the full finite-time reponse defined in Eqs.~\eqref{amp4} and \eqref{rate} in the numerical analysis below.

\subsection{Graphical results and analysis}

We now explore the consequences of the result \rf{rate}
using numerical evaluation.
Firstly, 
note the dependence of the result \rf{rate} on the dimensionless ratios of $m/a$ and $\Om/a$ (in natural units).
In fact, 
the only dimensionfull quantity left in Eq.~\rf{rate}
is the Gaussian width $\si$.
Our interest here is to compare the conventional behavior of this excitation 
rate with the case of $\de$, and $\ka$ differing from $0$ (and thus $\ze \neq 1$ as well).

\begin{center}
\begin{figure}[htp!]
\includegraphics[width=120mm]{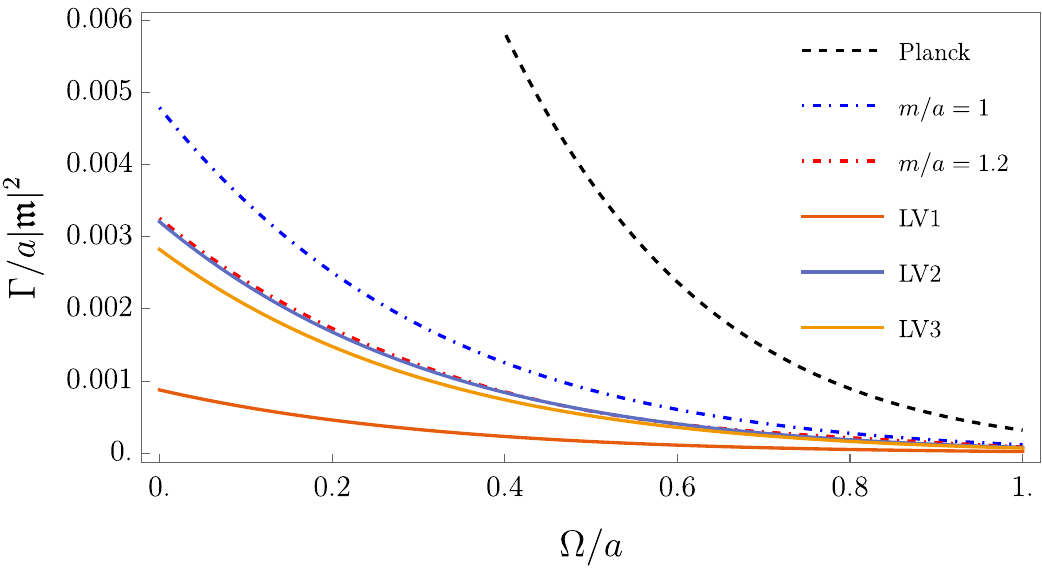}
 \caption{Numerical solutions of the excitation rate $\Ga$ in Eq.~\rf{rate} versus $\Om/a$ for different values of the coefficient-dependent quantities $\ze$, $\de$, and $\ka$ and sample values for the ratio $m/a$.  
 Note that the rate is scaled by $|\mathfrak{m}|^2$.}
 \label{gammaplot1}
\end{figure}
\end{center}

The average rate $\Ga$ is plotted in Figure \ref{gammaplot1}.
For ease of comparison
we have scaled the rate $\Ga$ by the factor $a|\mathfrak m|^2$ (thus we plot ${\cal F}_g/(\sqrt{\pi}a\si |\mathfrak m|^2 )$).
We include the Planck spectrum case and 
two sample values for the case of vanishing coefficients $k_\mn=0$ (conventional massive rate).
The first of the massive cases sets $m/a=1$, 
while the other mass value $m/a=6/5$.
One can see that smaller mass rate is steeper and approaches the Planck spectrum (the massless case).
On the other hand, 
increasing the mass to $m/a=6/5$ effectively reduces the excitation rate curve significantly.
The three cases for sample nonzero Lorentz violation 
are chosen as follows,
\begin{center}
{\bf LV1}: $\de=0.033$,\;$\ka=0.022$, \quad {\bf LV2}: $\de=6.7\times10^{-7}$, \;$\ka=0$, \quad {\bf LV3}: $\de=8.6\times10^{-6}$,\;$\ka=6.7\times10^{-6}$
\end{center}
all for fixed $m/a=1$.
For all of these cases, 
the effects are small enough that effectively, 
$\det K = -1$, 
$\ze =1$. 

We find that activating the coefficients for spacetime-symmetry breaking with coefficients with magnitudes much larger than listed above, brings to the rate to zero.
This qualitavely shows that small Lorentz violation can have a large effect on the (Unruh-deWitt) UdW signal.
The origin for this ``amplification" is the $\Si$-dependent terms in Eq.\ \rf{Qr}, 
which grows without limit with $\Si$ and hence dampens the rate significantly. 
This is because the $\Si$-dependent factor lies in the denominator and in the argument of the Bessel function.

\begin{center}
\begin{figure}[htp!]
\includegraphics[width=120mm]{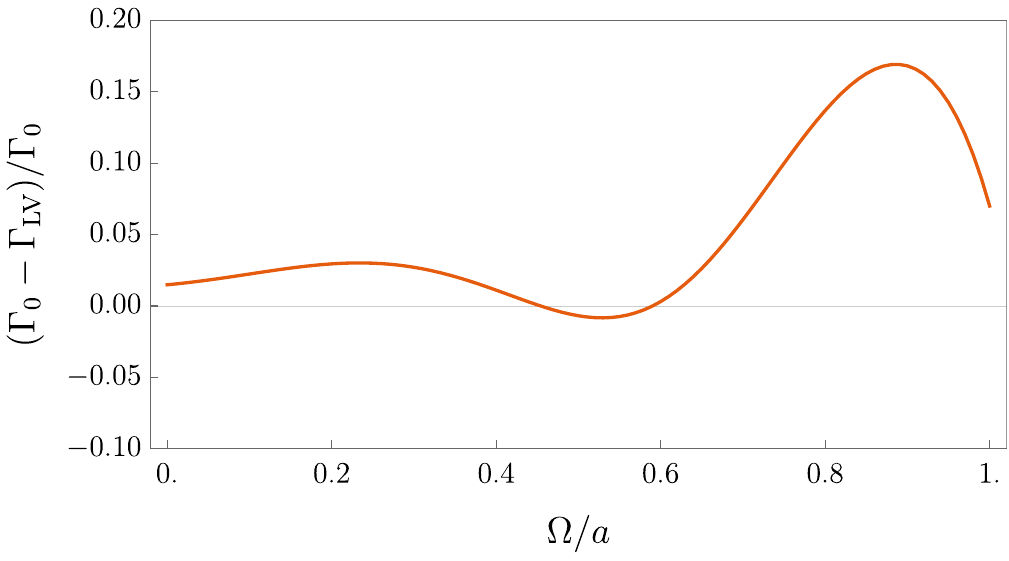}
\caption{Discrepancy between the curves in Figure \ref{gammaplot1}.}
\label{gammaplot2}
\end{figure}
\end{center}

A clear feature of the plot in figure \ref{gammaplot1} is
the similar appearance of some of the curves with nonzero $k_\mn$ (``LV" curves) 
and the conventional massive curves.
For instance this is the case with the $m/a =1.2$ curve and the LV case 2 curve.
One way to compare these curves is to subtract the curves that overlap.
We plot the discrepancy $(\Ga_0 - \Ga_{LV})/\Ga_0$ between the symmetry-breaking case and the conventional mass case in Figure \ref{gammaplot2}.

This last plot shows a clear difference between the functional forms for the results, 
indicating that the effects of Lorentz violation cannot merely be mimicked by mass alone.
Therefore it would be of interest experimentally to try to test for variations in the rate spectrum.
Whether experiments can do this is another matter.
Things are further complicated since scalar particles are not actually measured and the calculation would need to be generalized to photons and more realistic detectors
described by fermions.

\begin{center}
\begin{figure}[htp!]
\includegraphics[width=120mm]{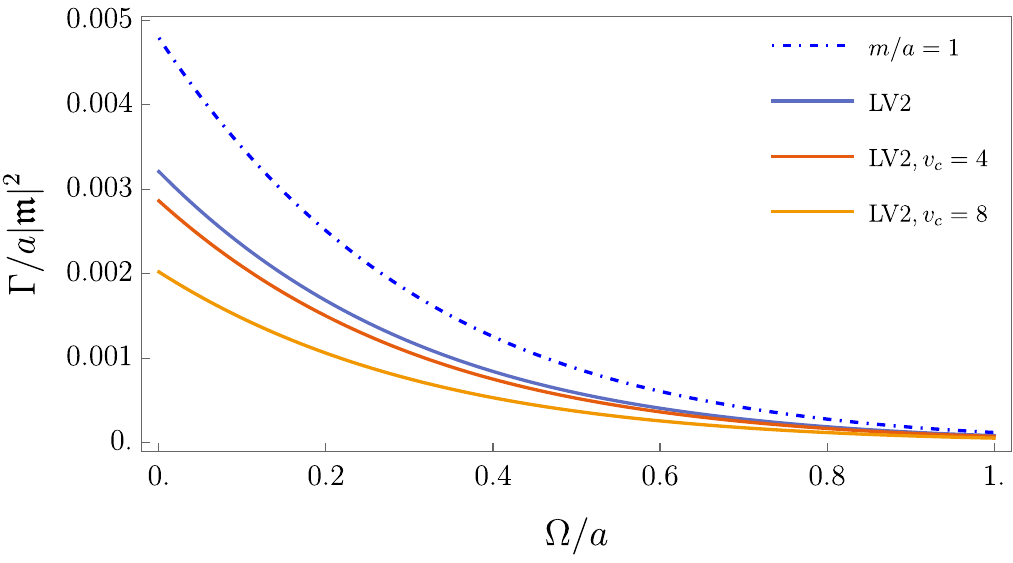}
\caption{Excitation rate for three example values of $v_c$, 
indicative of running the detector at different portions of its hyperbolic motion.}
\label{nonstation}
\end{figure}
\end{center}

There is another way, 
and a quite striking one,
in which the results with $k_\mn \neq 0$
differ from conventional ones.
As hinted at earlier, 
the results will depend on the measurement time window (non-stationarity).
To illustrate this, 
we include a plot where the center of the response window is varied away from $v_c=0$ (which refers to $\ta=0$, 
when the accelerated trajectory crossed the inertial $x$-axis).
In Figure \ref{nonstation}, 
we sample the excitation rate with $v_c=0,4,8$.
One can immediately see the nonstationarity of the result.  
It appears to matter when, 
relative to the accelerating trajectory,
the detector is running.

Taking these features together, 
we posit that this preliminary work represents a ``proof of principle" showing that minimal ``dimension 4" operators in the action for Lorentz violation could alter the thermal spectrum for an accelerating detector.\footnote{As the mechanism here affects the thermal spectrum, we note that a study of CPT and Lorentz violation in statistical mechanics was done in Ref.\ \cite{cm04}.}
Future studies are therefore in order to explore this idea with photons.
Note that much larger deviations may be expected for nonminimal coefficients for Lorentz violation (e.g., see the transition rate plots in Refs.\ \cite{Husain:2015tna, Louko:2017emx}).

\subsection{Application to experiment}

We discuss briefly the experimental implications of Eq.~\rf{rate}.
Clearly, 
an idea like using macrosopic detectors with large enough accelerations to measure the Unruh effect is impossible at present.
This can be seen from the Unruh temperature relation, 
in SI units, $T=(4 \times 10^{-21} K s^2 m^{-1}) a$.
On the other hand, 
subatomic particles have achieved high accelerations $\sim 10^{20} m s^{-2}$ but
the time scale is extremely short, 
and it is challenging to separate thermal radiation from larger conventional radiation effects of a charged particle.
Nonetheless, 
there are existing proposals that could bring detectibility within reach
\cite{lasers99,Bell:1986ir,schutzhold06,6z1l-kkmk,stargen22,Peng:2026bdm,Wu:2026phx}.
Further, 
some modeling using theoretically inferred Unruh radiation exists from experimental data \cite{Lynch:2019hmk}.
Nonetheless, 
we still discuss potential observation of spacetime-symmetry breaking effect in proposed tests.

First, 
we discuss an important imprint of hypothetical spacetime-symmetry breaking.
That is, 
that the coefficients $k_\mn$ will develop time dependence if they are re-expressed in terms of fixed Sun-centered Frame (SCF) coefficients.
The coefficients in the SCF are assumed to be constant on the time-scales of experiments and observations, as discussed elsewhere.\cite{kl99,km02,kt11,datatables}.
For instance, 
for the combinations of coefficients appearing in \rf{zedeka},
$k_{00}$ is a rotational scalar under observer transformations, 
while $k_{11}$ and $k_{01}$ change under rotations.
The standard slow motion coordinate transformation from the laboratory frame
(in which \rf{rate} is expressed) to the SCF involves a time-dependent Lorentz transformation
including a boost and rotation.
For simplicity we ignore the boost part since it is suppressed by the Earth's orbital velocity
$\be \approx 10^{-4}$.
The relavant lab frame coefficients expressed in terms of the SCF coefficients 
are, 
\beq
\bal
k_{11}^{\rm lab} &= k_{ZZ} \sin^2 \ch - \sin 2\ch (k_{XZ} \cos \om T + k_{YZ} \sin \om T)
\\
\quad 
&+\cos^2 \ch [ \frac 12 (k_{XX}+k_{YY}) + \frac 12 (k_{XX}-k_{YY}) \cos 2\om T  +k_{XY} \sin 2\om T ]
\\
k_{01}^{\rm lab}&= \cos\chi[k_{TX} \cos{\omega T} + k_{TY}  \sin \om T] - k_{TZ} \sin \ch,
\label{labtoscf}
\eal
\eeq
where the SCF coordinates are labeled with capital letters ($T$, $X$, $Y$, and $Z$), 
$\ch$ is the co-latitude of the experiment and $\om$ is the sidereal rotation frequency of the Earth.
If one looks for harmonic variations in experimental data, one accesses sensitivity in the transition rate measurement to multiple {\it a priori} independent coefficients.
Harmonic searches for Lorentz and CPT violation using this type of transformation abound 
in the literature (see a complete list in \cite{datatables}).
So to put it briefly, 
sidereal variation in the measurement of a signature of the Unruh effect could point to a spacetime symmetry breaking origin. We also note that the existence of preferred frames or prior geometry studied in this paper, which also appears in for example Ho\v{r}ava-Lifshitz gravity~\cite{Horava:2009uw} or related khronometric theories can leave distinctive experimental signatures in the Unruh effect when evaluated on different experimental trajectories in the form of anisotropies or sidereal variations. Although interesting, we do not pursue this further here.

It is useful give a crude estimate of the sensitivity of an experiment to the coefficients 
$k_\mn$, 
supposing that scalar radiation could be measured. 
We take as an example the experimental scenario described by
Ref.\ \cite{6z1l-kkmk}.
These authors describe an electromagnetic cavity test with collections of atoms accelerated.
Under the right conditions, 
like excellent mirrors with near perfect reflectivity to within $10^{-8}$, 
a ``superradiant burst" from the Unruh acceleration could be observed.
In fact, 
the effective radiation rate could be about 50 times a comparable inertial one, 
indicating a roughly $1/50$ sensitivity factor.
Figure \ref{gammaplot2} can be used to show the Lorentz-violating effect for coefficients of order $\de \sim 7 \times 10^{-7}$. 
The portion of the discrepancy curve from $\Om/a \sim 0.6 - 1.0$ has an average of about $0.10$.
Using $0.02$ as the measurement uncertainty for comparison, 
a signal like in Figure \ref{gammaplot2} could be discerned with some statistical significance.
Therefore we can say, 
crudely,
that coefficients of order $10^{-6}$ could be potentially ruled out with such an experiment. 
The reader is cautioned, 
however, 
that what one needs is 
a thorough analysis of the Unruh effect with elctromagnetic fields to obtain a proper estimate of sensivity to Lorentz-violation coefficients.

In a realistic particle experiment, 
one would need to completely describe the interactions within the full CPT and Lorentz-violating Quantum Electrodynamics (QED) framework 
\cite{ck98,datatables}.
This is a broad subset of the EFT framework by itself and has been studied
extensively in theory and experiment.
In fact, 
striking Lorentz-symmetry breaking effects have been pointed out in the past literature
\cite{Kostelecky:2002ue}.
The current limits to the photon sector coefficients in the minimal framework are quite stringent from electromagnetic cavity tests over the years
\cite{Muller:2007zz,Herrmann:2009zzb,datatables}.
The important thing here is that any ``extracted"  acceleration-only effects from the net observables must be done carefully.
Consideration must be made to any spacetime-symmetry breaking coefficients that play a role in all of the aspects of particle experiment observables.
We leave this broader program to future work.

\section{Lorentz violation in the detector trajectory}
\label{FTM}

In the preceding sections, we treated Lorentz violation in the scalar sector using the full quantum detector response. We now consider the complementary case in which the scalar field is conventional while the accelerated trajectory is modified. It is useful to compute the proper-time Fourier transform of a scalar plane wavce evaluated along the modified trajectory.
This method was adopted in,
for example, 
Refs.\ \cite{Alsing:2004ig} and \cite{kleinert2016particles}.

We consider the motion of the point particle under the effects of the $c_\mn$ coefficients in the action in equation \rf{particle}.  
To model a uniform acceleration we suppose that this point particle is charged and coupled to an electromagnetic field, described by the field strength tensor $F_\mn$.
Variation of the action \rf{particle}, augmented with a standard electromagnetic coupling, with respect to the trajectory $x^\mu (\la)$ yields the equations of motion: 
\beq
(\eta_{\mu\nu}+c_{\mu\nu})\frac{dU^\nu}{d\tau} = \frac qm F_\mn U^\nu,
\label{chargeEOM}
\eeq
where, after the variation, 
the parameter $\la$ has been chosen as the proper time $\ta$.
The latter is defined by:
\beq
d\ta^2 = ( \et_\mn + c_\mn )dx^\mu dx^\nu, 
\label{modtau}
\eeq
and thus $U^\mu = dx^\mu/d\ta$.

A critical constraint follows directly from Eq.~\rf{modtau}
upon dividing by differential coordinate time $dt$,
\beq
(\et_\mn + c_\mn) U^\mu U^\nu =-1,
\label{Uconstraint}
\eeq
which is a modification of the standard normalization 
for the four-velocity.
Note that this constraint is consistent with solutions to \rf{chargeEOM} which can be verified upon contraction of this equation with $U^\mu$.
This yields $(\et_\mn + c_\mn)U^\mu d U^\nu/d\ta=0$, 
which is of the form of a constraint but is not independent of \rf{Uconstraint}.

Another constraint is that for the four-acceleration $a^\mu = dU^\mu/d\ta$.
In the usual case, 
this quantity has fixed magnitude $a^\mu a_\mu = a^2$, 
where $a$ is the constant acceleration identified by accelerometers along the trajectory \cite{mtw}.
When $c_\mn \neq 0$, 
the constraint is modified to
\beq
(\et_\mn + c_\mn) a^\mu a^\nu = {\tilde a}^2,
\label{Aconstraint}
\eeq
where the tilde allows for a scaling of $a$, to be determined.

We assume that only a constant electric field exists ($F_{10}=E=-F_{01}$) in the chosen coordinates and all other components of $F_\mn$ are zero. 
A straightforward method to solve \rf{chargeEOM} is to treat it as a first order system as follows.
In matrix form, 
Eq.~\rf{chargeEOM} reads
\beq
\bal
P\frac{dU}{d\tau} &= \frac{q}{m}FU, \\
\frac{dU}{d\tau} &= \frac{q}{m}P^{-1}FU,
\label{matrixEOM}
\eal
\eeq
where in the second line we assume that $P_\mn = \et_\mn + c_\mn$
is invertible (e.g., see the conditions on $K^\mn$ in \rf{Kinv}).
In general this is a $4\times4$ matrix problem with four unknown functions $U^\mu$.
For simplicity in this presentaiton we assume only nonzero $c_\mn$ in the ``1+1" submatrix; 
only $c_{00}$, 
$c_{11}$, 
and $c_{01}$ are nonzero, 
and we assume $U^2=U^3=0$.
The problem is then reduced to a $2\times2$ matrix first order differential equation problem.

The general solution to this first order system of equations takes the form
\begin{equation}
    U = c_1 e_1 e^{\la_1 \ta} + c_2 e_2 e^{\la_2 \ta},
    \label{gensoln}
\end{equation}
where $\la_1$, $\la_2$ and $e_1$, $e_2$ are the eigenvalues and eigenvectors for the $2\times2$ matrix $\tilde P = (q/m) P^{-1}F$, 
and $c_1$ and $c_2$ are arbitrary constants.
Specifically, 
this matrix takes the form
\beq
\tilde{P} = \frac {a}{d_c}
\begin{bmatrix}
c_{01} & 1+c_{11} \\
1-c_{00} & -c_{01} \\
\end{bmatrix},
\label{2x2mat}
\eeq
where $d_c = c_{01}^2+(1-c_{00})(1+c_{11})$,
and $a=qE/m$.
The eigenvalues and eigenvectors of this matrix are found to be 
\beq
\bal
\la_1 = \tac, \quad
\la_2 = -\tac , \quad
e_1 = 
    \left[1 ,
    \frac{\sqrt{d_c} - c_{01}}{1+c_{11}}\right]^{\rm{T}}
, \quad e_2 = 
   \left[1,
    \frac{-(\sqrt{d_c} + c_{01})}{1+c_{11}}\right]^{\rm{T}}
\label{eigen}
\eal
\eeq
where $\tac= a /\sqrt{d_c}$.

Insertion of Eq.~\rf{eigen} into Eq.~\rf{gensoln} results in a general solution which is then subject to the constraint \rf{Uconstraint}
and initial conditions.
Inserting the general solution into Eq.~\rf{Uconstraint} yields an equation relating $c_1$ and $c_2$, thus eliminating one constant.
Next, 
the initial condition on $U^\mu$ at $\ta=0$ is chosen as $U^1(0)=0$, to eliminate the spatial vector piece of the four-velocity at this moment.
These conditions fix $U^\mu$ uniquely.
To obtain the trajectory $x^\mu (\ta)$, 
we simply integrate $U^\mu$ with respect to $\ta$, 
and impose the initial conditions $x^0(0)=t(0)=0$ (for synchonization) and $x^1(0)=x(0)=\sqrt{1-c_{00}}/a$.
The solutions for the four-velocity and trajectory are then given by
\begin{equation}
    U^\mu = 
    \left[\frac 1{\sqrt{1-c_{00}}}\left(\cosh(\tac \ta)+ \frac{c_{01}}{\sqrt{d_c}}\sinh( \tac \tau)\right), 
    \sqrt{\frac{1-c_{00}}{d_c}}\sinh( \tac \tau), 
    0, 
    0\right]^{\rm{T}},
\end{equation}
and
\begin{equation}
    x^\mu = 
   \frac{1}{a}\left[ \frac1{\sqrt{1-c_{00}}}\left(\sqrt{d_c} \sinh(\tac \tau)+c_{01} (\cosh(\tac \tau)-1)\right),
    \sqrt{1-c_{00}}\cosh(\tac \tau),
    0,
    0\right]^{\rm T}
     \label{modX}
\end{equation}
Note that Eq.~\rf{Aconstraint} is automatically satisfied at this point, 
as can be directly checked.

\begin{figure}[t]
  \centering
     \begin{subfigure}[t]{0.49\textwidth}
       \centering
         \includegraphics[width=\textwidth]{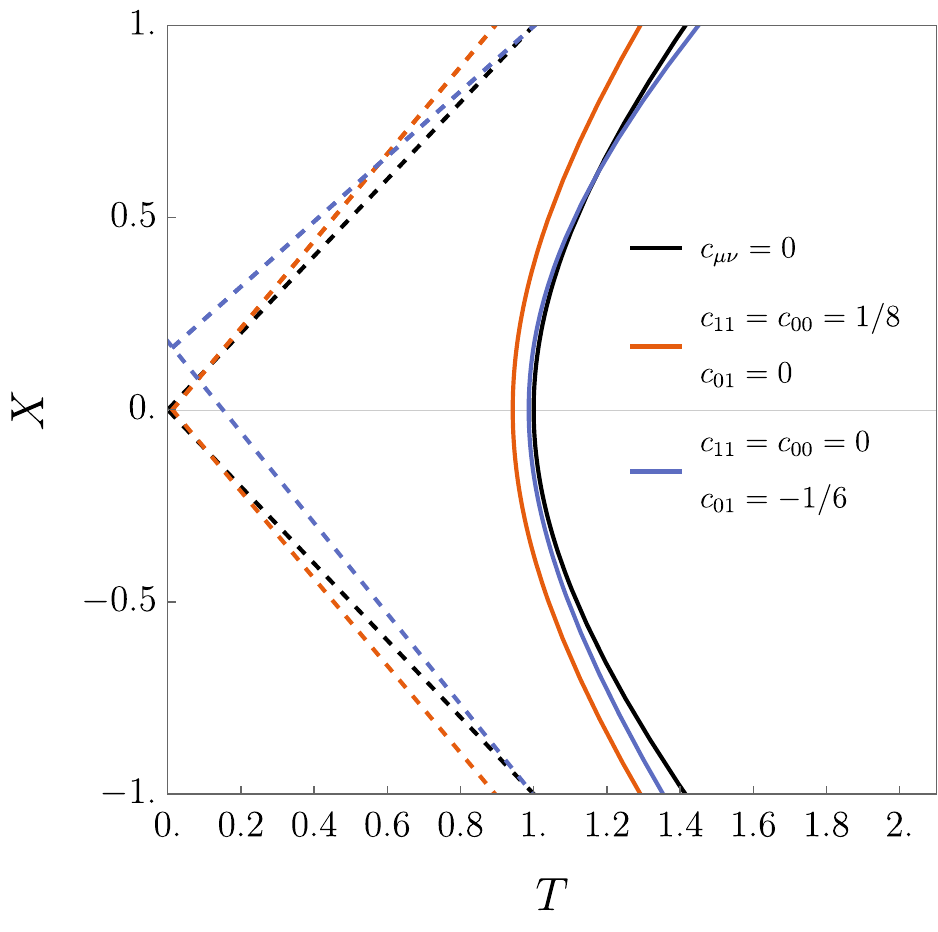}
         \caption{Trajectory described by Eq.~\rf{modX} for sample values of $c_{\mn}$.  The dashed line represent asymptotes of these trajectories.}
         \label{traj1}
     \end{subfigure}
     \hfill
     \begin{subfigure}[t]{0.46\textwidth}
         \centering
\includegraphics[width=\textwidth]{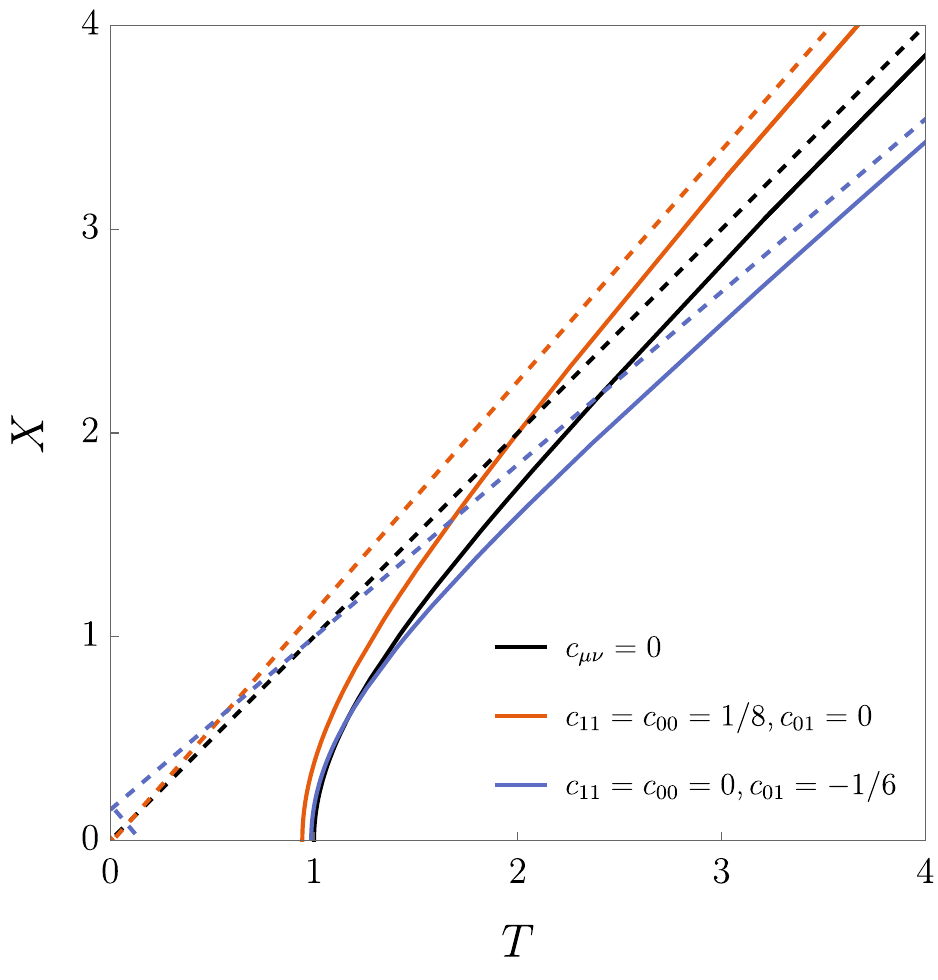}
         \caption{Asymptotic behavior of the trajectories of Figure~\ref{traj1}, 
         showing that they approach a skewed lightcone.}
         \label{traj2}
     \end{subfigure}
     \caption{Acceleratingtrajectory plots of Eq.~\rf{modX};
     the axes are related to $x$ and $t$ coordinates by $X=at/\sqrt{d_c}$, $T=ax/\sqrt{d_c}$.}
\end{figure}

To our knowledge, 
these modified accelerating trajectories have not been published before.
It is useful to visualize these solutions for special cases of the coefficients $c_\mn$.
In Figures~\ref{traj1}-\ref{traj2}, 
three sample trajectories are plotted.
Figure~\ref{traj1} shows the modified hyperbolic trajectories for two coefficient values compared with the standard case (black lines).
The dashed lines are the asymptotes of these trajectories.
The asymptotic behavior is shown in Figure~\ref{traj2}.
Note that the modified hyperbolas approach modified massless trajectories resembling the published plots in Ref.\ \cite{Bailey:2023lzy}.

\subsection{Fourier spectrum of a scalar wave for a modified accelerating observer} 
\label{Spectrum}

Now consider that the scalar field is unmodified ($k_\mn = 0)$.
The action \rf{scalar_act} then yields the usual Klein-Gordon field equations $(\Box - m^2)\vph=0$.
We consider a scalar plane wave propagating along some $\hat n$
direction.
\beq
\vph = A e^{i (\om t - p \hat n \cdot \bf r)},
\label{pw}
\eeq
where $\om = \sqrt{p^2+m^2}$, $p$ is the spatial momentum magnitude, $A$ is a constant amplitude, and $\bf r = (x,y,z)$.
Next we evaluate the wave along the trajectory of the accelerating observer in Eq.~\rf{modX}, whereupon Eq.~\rf{pw} becomes
\beq
\vph = A e^{i \ps (\ta)},
\label{pw2}
\eeq
where the phase $\ps (\ta)$ is given by
\beq
\ps = \frac {\om}{\sqrt{1-c_{00}} \, \tac} \sinh \tac \ta 
+ \left(  
\frac {\om c_{01}}{\sqrt{1-c_{00}} \, a} 
- \frac { \sqrt{1-c_{00}} } {a} p \cos \th 
\right) 
\cosh \tac \ta 
- \frac {\om \, c_{01}}{\sqrt{1-c_{00}} \, a },
\label{phase}
\eeq
where $\cos \th  =  \hat p \cdot \hat n$.
Note that the last term on the right side is $\ta$-independent.

To determine the spectrum of this plane wave, measured by an accelerating observer, 
we take the Fourrier transform of $\vph (\ta)$ with respect to $\ta$, using $\Om$ as the frequency.
This method is similar to that taken in Ref.\ \cite{Alsing:2004ig}.
Specifically,
we calculate
\beq
{\tilde \vph} (\Om) = \frac {1}{\sqrt{2\pi}} \int_{-\infty}^\infty A e^{i \ps(\ta)} e^{i \Om \ta} d\ta.
\label{ft1}
\eeq
To evaluate this integral it is convenient to re-write $\ps$ as,
\beq
\ps = {\cal A} e^{\tac \ta } 
+ {\cal B} e^{-\tac \ta } 
+ {\cal C},
\label{phase2}
\eeq
where the $\cal A$, $\cal B$, and $\cal C$ are $\ta$-independent quantities given by
\beq
\bal
{\cal A} &= \frac 12 \left(
\frac {\om}{\sqrt{ 1-c_{00} }a}
(c_{01} + \sqrt{d_c}) - \frac { \sqrt{ 1-c_{00} } }{a} p \cos \th \right),\\
{\cal B} &= \frac 12 \left(\frac {\om}{\sqrt{ 1-c_{00} }a}
(c_{01} - \sqrt{d_c}) - \frac { \sqrt{ 1-c_{00} } }{a} p \cos \th \right),\\
{\cal C} &= - \frac {\om \, c_{01}}{\sqrt{1-c_{00}} \, a }.
\eal
\label{abc}
\eeq
It is advantageous to change variables from $\ta$ to $y$, via $y=e^{\tac \ta}$.
This changes the integral in Eq.~\rf{ft1} to
\beq
{\tilde \vph} = A \frac {e^{i{\cal C}} }{\sqrt{2\pi} \tac} \int_0^{\infty} e^{i \left( {\cal A} y + \frac {{\cal B}}{y} \right) } y^{\frac {i \Om}{\tac} -1 } 
dy,
\label{ft2}
\eeq
which can be evaluated using a modified Bessel function of the second kind $K_n (z)$, 
if $\cal A$ and $\cal B$ are given small positive imaginary parts ($+i \ep$).
The result is  
\beq
{\tilde \vph} = \frac {e^{i{\cal C}} }{\sqrt{2\pi} \tac} 2 
\left( \frac {{\cal B}+i\ep}{{\cal A}+i\ep}\right)^{i\frac {\Om}{\tac} }
K_{-i\frac {\Om}{\tac}} (-2i \sqrt{
({\cal A}+i\ep)({\cal B}+i\ep ) } ).  
\label{ft3}
\eeq

The desired result is a picture of the spectrum detected by the accelerating observer moving along the modified trajectory \rf{modX}.
The modulus of the Fourier transform \rf{ft3}, 
upon inserting the quantities in Eq.~\rf{abc}, 
can be written as,
\beq
|{\tilde \vph}|^2 = \frac {2 A^2}{\pi \tac^2} 
|K_{-i\frac {\Om}{\tac}} 
(2 \sqrt{ \al + i \ep })|^2.
\label{ftsq}
\eeq
The quantity ${\al}$ is given by
\beq
\al = \frac {1}{4\tac^2} 
[ m^2 (1+c_{11}) + p^2 (\sin^2 \th + c_{11} + c_{00} \cos^2 \th ) 
+ 2 c_{01} \sqrt{p^2+m^2} \, p \cos \th ]
\label{cal}.
\eeq
Apart from scalings, 
the coefficients modify the argument of the Bessel function in a orientation dependent way due to the presence of $\th$.
Note that even in the conventional case, 
the quantity in brackets becomes
$m^2+ p^2 \sin^2 \th$, 
which retains a dependence on the direction of the plane wave relative to the accelerated trajectory.
If the plane wave is lined up so that $\th=0$, 
the the argument of the Bessel function only depends on the ratio $m/a$ in the conventional case.

\begin{figure}[t]
    \centering
\includegraphics[width=\linewidth]{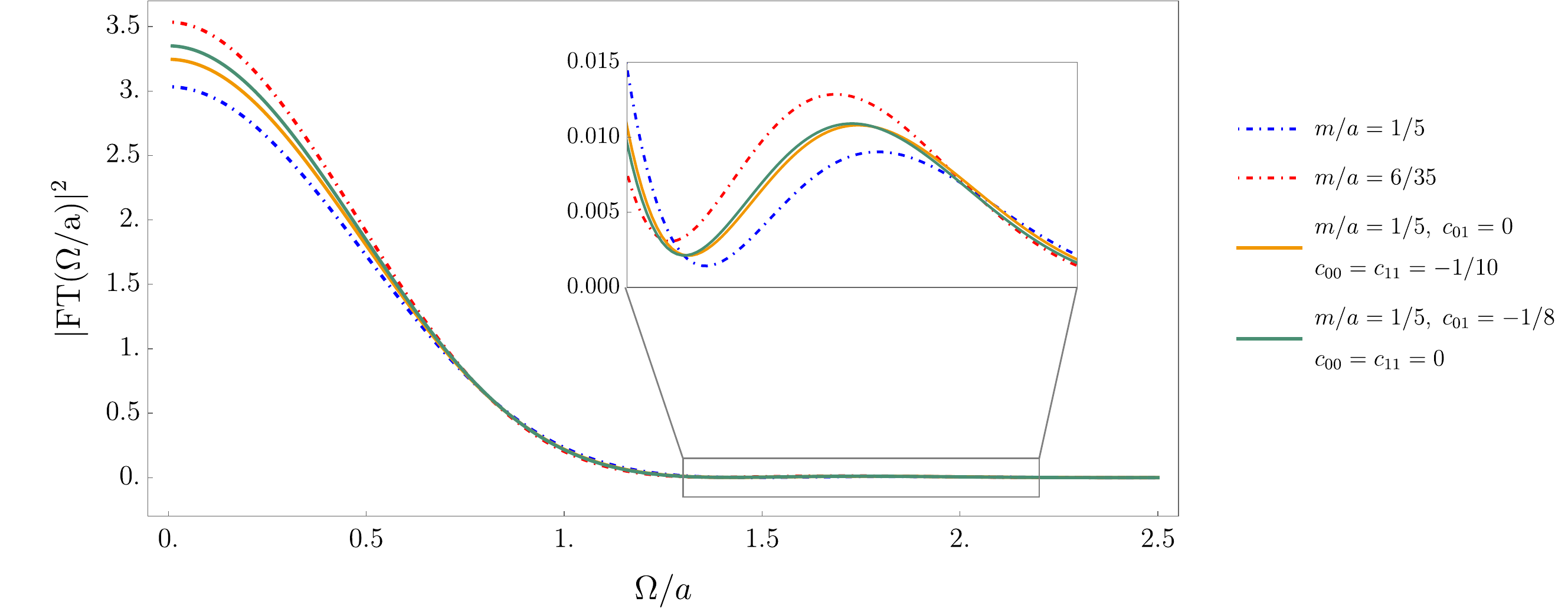}
    \caption{The Fourier transform in Eq.~\rf{ftsq};
     for these plots the axes are related to $x$ and $t$ by $X=at/\sqrt{d_c}$, $T=ax/\sqrt{d_c}$.}
    \label{fig:ftplot}
\end{figure}

To get an idea of the types of effects of the terms in $\al$ on the spectrum, we plot it in Figure~\ref{fig:ftplot}.
In the figure, 
it can be observed that the symmetry breaking cases (with nonzero $c_\mn$ coefficients) are behaving qualitatively like an effective mass shift. 
The two cases with distinct coefficients can be ``trapped" between the two (conventional) curves for nearby mass values.
Also, 
this behavior can be seen for the small peak around $\Om/a \approx 1.8$ in the inset of Figure~\ref{fig:ftplot}.
It is challenging to discern the difference between the effect of the coefficients and adjusting value of the mass $m$.
Thus it is not clear if there is a striking difference 
that could be gleaned from experiment.
Nonetheless, 
the $c_{01}$ coefficient in \rf{cal} shows a distinct dependence on $p$ that does not occur for the conventional case.
In addition, 
the coefficient time-dependence considerations of the previous sections also apply here and the signal in the laboratory can be modulated by the Earth's sidereal rotation frequency.
We leave it as an open problem to study possible tests with this alternate scenario.

\section{Discussion \& conclusions}
\label{sec:conclusions}

In this paper, 
we studied the Fulling-Davies-Unruh effect in the effective-field theory framework for tests of Lorentz violation introduced in Section~\ref{eft}. Most of the analysis was done with the coefficients for Lorentz violation in the scalar-field sector, while the detector follows the usual uniformly accelerated trajectory. We also considered a complementary case where the scalar field is conventional and the symmetry-breaking coefficients appear in the point-particle sector. In this way, the effects coming from the field correlations and those coming from the accelerated trajectory can be considered separately.

For the scalar field, in Section \ref{sec:scalarfield_quantisation}, 
we carried out canonical quantization without expanding to leading order in the coefficients $k_\mn$. 
The dispersion relation is given in Eq.~\rf{disp}, 
the quantized field and momenta
are given in Eqs.\ \rf{fieldQ},
and the exact massless and massive Wightman functions are given in Eqs.~\rf{eq:general_massless_Wightman} and \rf{twopointm3}. 
An advantage of this approach is that the dispersion relation, 
the vacuum prescription, and the two-point function all follow from the same local action in Eq.~\rf{scalar_act}. 
This is somewhat different from a number of treatments in the literature where Lorentz violation is introduced through a modified dispersion relation, polymer quantization, 
or a related modification of the field theory \cite{Agullo:2008qb,Campo:2010fz,Hossain:2015xqa,Husain:2015tna,Davies:2023zfq}.

In Section \ref{sec:UnruDavies_boboliubov},
one of the main results from the field analysis concerning boost symmetry was established. 
The most general constant $K^\mn$ which preserves the boost symmetry in the $x^0$--$x^1$ plane is given in Eq.~\rf{eq:general_boost_invariant_K}. 
Note that this does not require the theory to be fully Lorentz invariant, since the transverse part of $K^\mn$ can still be anisotropic. Also, the condition for stationarity along a single detector trajectory is weaker, and is given in Eq.~\rf{eq:worldline_stationarity}. Thus Lorentz violation by itself does not necessarily imply that the accelerated response is nonstationary.
We may use this result when comparing our work to that in the literature. 
It has been shown that the KMS property can be lost for preferred-frame modified dispersion relations and in polymer quantization \cite{Campo:2010fz,Hossain:2015xqa}. We find the same general behavior when the coefficients break the boost symmetry relevant to the accelerated observer. 
On the other hand, there are special Lorentz-violating choices for which this boost symmetry survives. This is also qualitatively consistent with the Lorentz-violating gravity case studied in Ref.~\cite{DelPorro:2023knh}, where an accelerated configuration with Unruh properties can still be defined. One should also keep in mind that loss of the KMS property does not necessarily eliminate every possible notion of a thermal detector response \cite{Carballo-Rubio:2018zll}, but generic boost breaking does removes the usual stationary Rindler description.

When the stationarity conditions are not satisfied, the finite-time detector response becomes relevant, and is given generically in Eq.~\rf{eq:switched_response2} in Section \ref{sec:finitetimeresponse}. 
The Lorentz-violating part depends on the average proper time $\Si=(\ta+\ta^\prime)/2$ through $\mathcal X(\Si)$. For the massless field, Eq.~\rf{eq:massless_finite_response} shows that this appears as a time-dependent scaling of the conventional result. In the massive case there is an additional effect, since Eq.~\rf{eq:massive_finite_response} has an effective ``mass'' term of the form
$m\rightarrow m\sqrt{\mathcal X(\Si)}$.
Thus the massive result has both a time-dependent prefactor and an effective time-dependent mass.
When boosts are broken, the center of the measurement window $\tau_c$ also matters, as can be seen directly in Eqs.~\rf{eq:exact_Gaussian_response} and \rf{eq:massive_small_breaking_response}. At leading order, the combination $H_{00}+H_{11}$ contributes to the part which is even under $\ta_c\rightarrow-\ta_c$, while the $H_{01}$ contribution is odd. In particular, the first-order $H_{01}$ contribution vanishes when the detector response is centered at $\ta_c=0$. Finite switching effects also occur in the conventional Lorentz-invariant detector problem \cite{Sriramkumar:1996finite,Louko:2008transition,Dickinson_2025,Stargen:2026finite}. The difference here is that there is an additional dependence on when the detector is ``on'' relative to the inertial frame time. Related issues involving finite-time and dynamical descriptions of the Unruh effect have also been discussed recently \cite{Stargen:2025thermality,Saha:2026prethermal}. There is also related work done from the point of view of modified dispersion relations \cite{Husain:2015tna,Davies:2023zfq,Louko:2017emx,Xu:2025smb}, although the origin of the time dependence here is directly connected with the coefficients in the local EFT.

A key result in the paper is the numerical study of the UdW detector amplitude and excitation rate, 
which is the subject of Section \ref{detector}.
The main equation for the amplitude $\cal F$ for a transition of the UdW detector from a ground state to an excited state, 
set up for numerical integration,
is given in \rf{amp4}.
We show the general behaviour of these effects using special cases which we solved numerically in Figs.~\ref{gammaplot1}-\ref{nonstation}. Figure~\ref{gammaplot1} shows the excitation measure in Eq.~\rf{rate} for several choices of the Lorentz-violating coefficients. Some of the curves look quite similar to the conventional massive result with a slightly different value of $m/a$. However, the discrepancy shown in Figure~\ref{gammaplot2} has a different functional form, which is consistent with our analytical result that the Lorentz-violating effect cannot in general be reproduced just by changing the mass.
Another difference can be seen in Figure~\ref{nonstation}, namely that keeping the coefficients fixed and changing the center of the Gaussian window changes the excitation curve. This gives a direct numerical example of the nonstationarity discussed above. For the choices used in the plots, noticeable changes occur for coefficient combinations roughly in the $10^{-7}$--$10^{-6}$ range, 
which gives an indication of the possible sensitivity of this type of observable. Small Lorentz-violating effects in accelerated-detector spectra have also been discussed in modified-dispersion calculations \cite{Husain:2015tna}, while momentum-resolved response \cite{Xu:2025smb} and detector coherence \cite{Wu:2026phx} have been considered as other possible probes.
Note also that the hyperbolic dependence on $\Si$ can make a small coefficient have a relatively large effect on the detector response, which is visible in the numerical results. At the same time, this means that a small coefficient does not by itself guarantee that the weak boost-breaking expansion remains valid for arbitrarily long measurements. The condition $|\delta\mathcal X/\mathcal X_0|\ll1$ must hold over the part of the trajectory where the switching function is appreciable.

There is also an experimental signature coming from the orientation of the laboratory. Equation~\rf{labtoscf} shows that coefficients which are constant in the standard Sun-centered frame become time dependent in the laboratory frame, with harmonics of the Earth's sidereal rotation frequency. Searches for this type of sidereal variation are common in tests of Lorentz and CPT symmetry \cite{kl99,datatables}. If an acceleration-related signal could be measured, a sidereal dependence would be one way to distinguish a spacetime-symmetry breaking effect from an ordinary change in the detector parameters. The sensitivity estimate made in Section~\ref{detector} is only preliminary, however, since the model considered here uses a scalar field rather than the electromagnetic and matter fields relevant for a realistic experiment.
We also considered the alternative case where the Lorentz violation is placed in the point-particle sector
in Section \rf{FTM}. 
For a constant electric field, the equations of motion can be solved exactly in the $1+1$ coefficient sector with $c_{00}$, $c_{11}$, and $c_{01}$. The resulting trajectory is given in Eq.~\rf{modX}. As shown in Figures~\ref{traj1} and \ref{traj2}, the trajectory is still hyperbolic in character, but the acceleration scale and the asymptotic directions are modified. In particular, $c_{01}$ produces a skewing of the trajectory. Lorentz-violating point-particle kinematics has been studied previously in a number of contexts \cite{kl10,kt11,Russell:2015finsler,Schreck:2015kinematics}, and the asymptotic behavior here also resembles the modified characteristic cones found for Lorentz-violating classical wave propagation \cite{Bailey:2023lzy}.
As a simple way to inspect the effect of this trajectory, we evaluated a conventional scalar plane wave along Eq.~\rf{modX} and took its Fourier transform with respect to the particle proper time. The result is given in Eqs.~\rf{ftsq} and \rf{cal}, with examples shown in Fig.~\ref{fig:ftplot}. Much of the change again resembles an effective shift in the mass. On the other hand, the $c_{01}$ term in Eq.~\rf{cal} has a distinct dependence on the momentum and orientation of the scalar wave. The proper-time Fourier-transform method has also been used for the conventional accelerated spectrum \cite{Alsing:2004ig,kleinert2016particles}.
It is important to note that the two ways of assigning the coefficients are not physically independent in all cases. As discussed in Section~\ref{eft}, coordinate redefinitions can move coefficients between the matter and field sectors for a single species \cite{Bailey:2004na,kt11,Yoder:2012ks,Bailey:2023lzy}. The measurable quantities therefore involve differences between sectors. Studying the two cases separately is nevertheless useful, since it makes it clear whether a given effect comes from the field correlations or from the detector trajectory.

There are several natural extensions of this work. The most important would be to repeat the calculation in full Lorentz-violating QED, where the accelerated particle, the electromagnetic field, and the detector response can be treated on the same footing. The photon sector is particularly interesting since Lorentz violation can produce birefringence and anisotropic propagation \cite{km02,km09,Nilsson:2023sxz}. It would be useful to determine how these effects enter an acceleration experiment. Higher-order terms in the EFT expansion are also of interest, since their dimensionful coefficients introduce additional powers of momentum and frequency, and previous detector calculations suggest that nonminimal terms can produce substantially different effects \cite{Husain:2015tna,Louko:2017emx}.

In this paper, we have studied the effect of minimal Lorentz violation from an EFT framework on the classic FDU effect for an accelerating detector.
The relation between the accelerated trajectory and the effective geometry of the field is important, 
and if the relevant boost symmetry is preserved, 
the usual stationary description can still be used. 
For more general coefficients, 
the response is nonstationary and depends on the finite measurement window, 
as encapsulated by the UdW detector amplitude obtained in this work.
The numerical examples show that this time dependence, together with changes in the frequency dependence of the excitation rate, 
can provide signatures that are different from a simple mass shift. These results suggest that accelerated systems may provide another way to probe Lorentz violation, although a realistic analysis will require extending the present scalar model to the matter and photon sectors.

\acknowledgements{N.A.N. was supported by the Institute for Basic Science under the project code IBS-49 R018-D3 and acknowledges support from PSL/Observatoire de Paris.
S.S. was supported by the Undergraduate Research Institute of Embry-Riddle Aeronautical University, Prescott and the Arizona NASA Space Grant consortium while working on this project.
Q.G.B.\ acknowledges Alexander S.\ Gard for work on related investigations of modified hyperbolic motion. OpenAI ChatGPT 6 Astra was used for literature search and numerical plots. All results were independently checked by the authors who take full responsibility for the scientific content of the manuscript.
}

\appendix

\section{Pullback of the detector response}
\label{app:pullbackiepsilon}
We derive the regulated effective interval used in the finite-time detector response of Section~\ref{sec:finitetimeresponse}. We begin with the preferred-frame positive-frequency prescription introduced in Section~\ref{wightman functions} and evaluate it on the standard uniformly accelerated trajectory.
We use the regulator defined in the preferred frame in Eq.~\eqref{eq:preferred_frame_iepsilon} as the displacement $\tilde{z}^0 \, \to \, \tilde{z}^0 - i\epsilon$, and writing the interval in the new coordinates as
\begin{equation}
    Q_{K,\epsilon}(z)=\mathbf{\tilde{z}}^2-(\tilde{z}^0-i\epsilon)^2,
\end{equation}
we see that this is equivalent to the same expression in the original coordinates
\begin{equation}
    Q_{K,\epsilon}(z)=H_{\mu\nu}\Delta x^\mu\Delta x^\nu+\frac{2i\epsilon}{\sqrt{\ga}}z^0+\epsilon^2,
\end{equation}
where $\ga\equiv-K^{00}>0$. This form is useful because the accelerated trajectory is specified in the original Cartesian coordinates. The coordinate separation on the accelerated worldline reads
\begin{equation}\label{eq:coordinate_separation_accelerated_worldline}
\Delta x^0=\frac{2}{a}\cosh{a\Sigma}\sinh{\frac{a\Delta}{2}
},\quad
\Delta x^1=\frac{2}{a}\sinh{a\Sigma}\sinh{\frac{a\Delta}{2}},
\end{equation}
and given that the interval defined by the effective metric $H_{\mu\nu}$ reads $Q_K(\Sigma, \Delta)=H_{\mu\nu}\Delta x^\mu\Delta x^\nu$, we have
\begin{equation}
    Q_K(\Sigma,\Delta)=-\frac{4\mathcal{X}(\Sigma)}{a^2}\sinh^2{\frac{a\Delta}{2}},
    \label{Qch}
\end{equation}
where
\begin{equation}
    \mathcal{X}(\Sigma)=-\frac12(H_{00} + H_{11})\cosh{2a\Sigma}-H_{01}\sinh{2a\Sigma}-\frac12
    (H_{00}-H_{11}),
    \label{chH2}
\end{equation}
as is shown in Eq.~\eqref{chH}.  
Under the positive-frequency $i\epsilon$ prescription, this gives us
\begin{equation}
     Q_{K,\epsilon}(\Sigma,\Delta)=-\frac{4\mathcal{X}(\Sigma)}{a^2}\sinh^2{\frac{a\Delta}{2}}+\frac{4i\epsilon}{a\sqrt{\ga}}\cosh{a\Sigma}\sinh{\frac{a\Delta}{2}}+\epsilon^2.
\end{equation}
Since the trajectory is future directed with respect to the preferred time orientation and $\mathcal{X}(\Sigma)>0$, we have that
\begin{equation}
    \lim_{\epsilon\to 0^+} Q_{K,\epsilon}(\Sigma,\Delta)=-\frac{4\mathcal{X}(\Sigma)}{a^2}\sinh^2{[\frac{a}{2}(\Delta-i0)]},
\end{equation}
in a distributional sense, which is the limit in which the positive-frequence Wightman function is defined. This is in fact the standard results multiplied by $\mathcal{X}(\Sigma)$, as expected. This establishes the regulated pullback used in Eqs.~\eqref{eq:QK_boost_breaking} and \eqref{eq:QK_Lambda} without assumptions of boost symmetry.

\section{Relations between $K^\mn$ and the matrix $M$}
\label{relations}

There are several relations between 
the inverse of $K^\mn$, 
called $H_\mn$, 
and the $3 \times3$ matrix $M$.
The kinetic tensor $K^\mn = \et^\mn + k^\mn$ was defined in Sec.\ \ref{eft}.
The inverse $H_\mn$ satisfies 
\beq
H_\mn K^{\nu \la} = \de^\la_{\pt{\la}\mu}.
\label{Hdef}
\eeq
On the other hand, 
the matrix $M$ was defined in Section \ref{sec:scalarfield_quantisation} by 
\beq
M^T \cdot (I+C) \cdot M=I
\label{Mdef}
\eeq
where 
\beq
C_{ij} = k_{ij} - \frac{k_{0i} k_{0j}}{K^{00}}.
\label{Cdef}
\eeq
Multiplying \rf{Mdef} from the right side with $M^{-1}$ and from the left side with $M$ shows that
\beq
M.M^T = (I+C)^{-1}.
\label{mmt}
\eeq

We now express $K^\mn$ in a suggestive space and time block matrix,
\beq
K^\mn = \begin{bmatrix}
K^{00} & K^{0j} \\
K^{0i} & K^{ij} \\
\end{bmatrix}.
\label{Kmat}
\eeq
This form is reminiscent of the $3+1$ form for the spacetime metric and its inverse in General Relativity \cite{Arnowitt:1959ah,mtw}.
With the appropriate definitions we can use this known form and its inverse to show that the inverse of \rf{Kmat} can be written as 
\beq
H_\mn = \begin{bmatrix}
\frac{1}{K^{00}} + \frac {K^{0l} H_{lm} K^{0m}}{(K^{00})^2} & 
-H_{jl} \frac {K^{0l}}{K^{00}} \\
-H_{il} \frac {K^{0l}}{K^{00}} & H_{ij}
\end{bmatrix},
\label{Hmat}
\eeq
where $H_{ij} = (M.M^T)_{ij}$.

\bibliography{refs}

\end{document}